\documentclass[prl,aps,amssymb,showpacs,footinbib,twocolumn,superscriptaddress,times,10pt, english]{revtex4-2}

\usepackage[english]{babel}
\usepackage[pdftex,colorlinks=true,linkcolor=blue,citecolor=blue,urlcolor=blue]{hyperref} 
\usepackage{amsmath,amssymb,amsthm}

\usepackage{amsmath,amssymb,mathrsfs}
\usepackage{latexsym}
\usepackage{graphicx}
\usepackage{epstopdf}
\usepackage{color}
\usepackage{amsfonts}
\usepackage{bm}
\usepackage{bbm}
\usepackage{multirow}
\usepackage{subfiles}

\usepackage{natbib}

\usepackage{ulem} 

\usepackage{tikz} 
\usepackage{tikz-feynman}

\usepackage{mathtools} 
\usepackage{physics}
\usepackage[version=3]{mhchem}

\usepackage{microtype} 
\usepackage{csquotes} 
\usepackage{booktabs} 

\makeatletter
\DeclareRobustCommand{\rchi}{{\mathpalette\irchi\relax}}
\newcommand{\irchi}[2]{\raisebox{\depth}{$#1\chi$}}
\renewcommand\@make@capt@title[2]{%
	\@ifx@empty\float@link{\@firstofone}{\expandafter\href\expandafter{\float@link}}%
	{\textbf{#1}}\@caption@fignum@sep#2\quad
}%
\makeatother

\makeatletter
\g@addto@macro\bfseries{\boldmath} 
\makeatother

\usepackage{graphicx}
\usepackage{placeins}

\usepackage{siunitx} 
\DeclareSIUnit\year{a}
\DeclareSIUnit\pixel{px}
\DeclareSIUnit\line{line}

\usepackage{scrlayer-scrpage} 
\usepackage{lastpage} 
\usepackage{placeins}

\graphicspath{{/}{figures/}}

\def\ie{\textit{i.e.\ },\ }

\usepackage{float}

\newcommand\boldvector[1]{%
  \ifcat\noexpand#1\relax 
    \boldsymbol{#1}
  \else
    \mathbf{#1}
  \fi
}

\def\ie{{\it i.e.\/}}

\newcommand{\supplement}[1]{%
  \clearpage%
  \title{#1}%
  \maketitle%
  \setcounter{equation}{0}%
  \setcounter{figure}{0}%
  \setcounter{table}{0}%
  \setcounter{page}{1}%
  \makeatletter%
  \renewcommand{\thesection}{S\arabic{section}}%
  \renewcommand{\thesubsection}{\Alph{subsection}}%
  \renewcommand{\theequation}{S\arabic{equation}}%
  \renewcommand{\thefigure}{S\arabic{figure}}%
  \renewcommand{\thetable}{S\Roman{table}}%
  \renewcommand{\thepage}{S\arabic{page}}%
  \makeatother%
}

\makeatletter
\def\maketitle{
\@author@finish
\title@column\titleblock@produce
\suppressfloats[t]}
\makeatother

\begin{document}

\title{Van-Hove tuning of Fermi surface instabilities through compensated metallicity}
\date{\today}

\author{Hendrik Hohmann}
\affiliation{Institute for Theoretical Physics and Astrophysics, University of W\"urzburg, D-97074 W\"urzburg, Germany}
\affiliation{W\"urzburg-Dresden Cluster of Excellence ct.qmat, Germany}

\author{Matteo D\"urrnagel}
\affiliation{Institute for Theoretical Physics and Astrophysics, University of W\"urzburg, D-97074 W\"urzburg, Germany}
\affiliation{W\"urzburg-Dresden Cluster of Excellence ct.qmat, Germany}
\affiliation{Institute for Theoretical Physics, ETH Z\"{u}rich, 8093 Z\"{u}rich, Switzerland}

\author{Matthew Bunney}
\affiliation{School of Physics, University of Melbourne, Parkville,
    VIC 3010, Australia}
\affiliation{Institute for Theoretical Solid State Physics,
    RWTH Aachen University, 52062 Aachen, Germany}

\author{Stefan Enzner}
\affiliation{Institute for Theoretical Physics and Astrophysics, University of W\"urzburg, D-97074 W\"urzburg, Germany}
\affiliation{W\"urzburg-Dresden Cluster of Excellence ct.qmat, Germany}

\author{\mbox{Tilman Schwemmer}}
\affiliation{Institute for Theoretical Physics and Astrophysics, University of W\"urzburg, D-97074 W\"urzburg, Germany}

\author{Titus Neupert}
\affiliation{Department of Physics, University of Z\"urich, Winterthurerstrasse
190, Zurich, Switzerland}

\author{\mbox{Giorgio Sangiovanni}}
\affiliation{Institute for Theoretical Physics and Astrophysics, University of W\"urzburg, D-97074 W\"urzburg, Germany}
\affiliation{W\"urzburg-Dresden Cluster of Excellence ct.qmat, Germany}

\author{Stephan Rachel}
\affiliation{W\"urzburg-Dresden Cluster of Excellence ct.qmat, Germany}
\affiliation{School of Physics, University of Melbourne, Parkville,
    VIC 3010, Australia}

\author{Ronny Thomale}
\email{Corresponding author: rthomale@physik.uni-wuerzburg.de}
\affiliation{Institute for Theoretical Physics and Astrophysics, University of W\"urzburg, D-97074 W\"urzburg, Germany}
\affiliation{W\"urzburg-Dresden Cluster of Excellence ct.qmat, Germany}


\begin{abstract}
\noindent Van-Hove (vH) singularities in the vicinity of the Fermi level
facilitate the emergence of electronically mediated Fermi surface
instabilities. This is because they provide a momentum-localized
enhancement of density of states promoting selective
electronic scattering channels. High-temperature
topological superconductivity has been argued for in graphene at vH filling which, however, has so far proven 
inaccessible due to the demanded large doping from
pristine half filling. We propose compensated metallicity as a path to
unlock vH-driven pairing close to half filling in an electronic
honeycomb lattice model. Enabled by an emergent multi-pocket
    fermiology, charge compensation is realized by strong breaking of chiral
    symmetry from intra-sublattice hybridization, while retaining vH dominated
    physics at the Fermi level. We conclude by proposing tangible realizations through quantum material design. 
    

\end{abstract}

\maketitle

\begin{figure*}
    \centering
    \includegraphics[width = 0.97\textwidth]{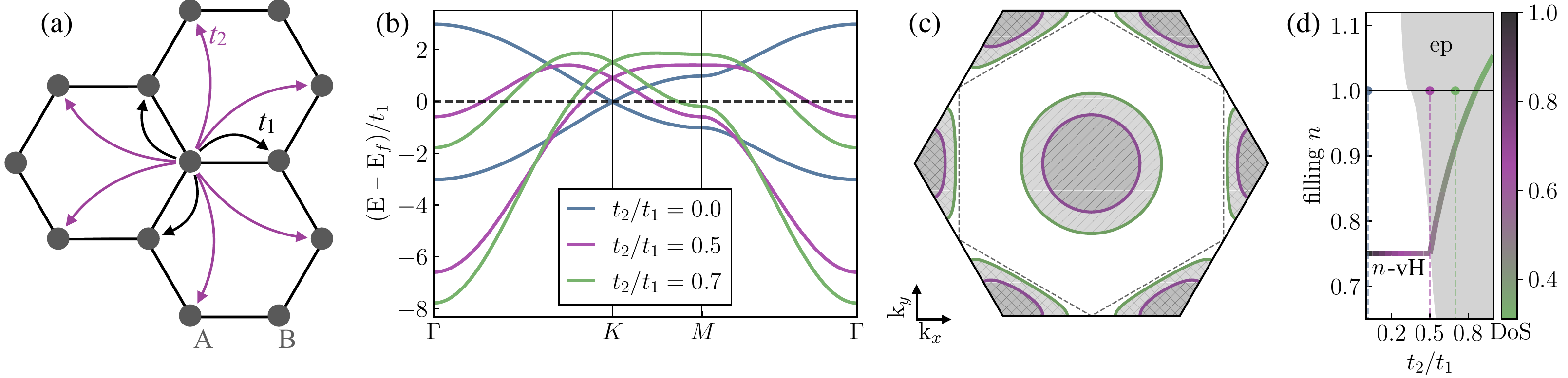}
    \caption{\textit{Absence of chiral symmetry and compensated
  metallicity}
    (a) Real space lattice structure and hybridization elements of the
    honeycomb Hubbard model, as given in Eq.~\ref{eqn:model}. (b) Single
    particle dispersion of the kinetic terms of Eq.~\ref{eqn:model}. Finite
    next-nearest neighbor hybridization breaks particle-hole symmetry and
    shifts the lower van-Hove singularity towards the Fermi level at half filling (dashed line).
    (c) Fermi surfaces at half filling within the hexagonal BZ, corresponding to the $t_2/t_1$ hopping values shown in (b). Electron (hole) pockets are indicated by crossed (diagonal) hatching. The perfectly nested outer FS is indicated by dashed line.
    (d) Lower vH filling and associated DoS at the Fermi level as a function of
    $t_2/t_1$. The dashed lines indicate $t_2/t_1$ values of (b) and
    (c). The opening of the electron pocket (ep) around $\Gamma$ (gray) shifts
    the vH towards pristine half filling ($n=1$). The numeric density of states
    (DOS) at an interval around vH is indicated along the vH line.
}
    \label{fig:bandstructure}
\end{figure*}

\paragraph{Introduction}

At sufficiently low temperature, metals as interacting many-body
electron systems tend to seek a
minimization of ground state energy through spontaneous symmetry
breaking. Starting from a Fermi surface (FS) configuration, the metal can
become unstable upon the formation of electronic order related to
magnetism, charge order, superconductivity, nematicity, and
more. Already from earliest descriptions of electronic collective
phenomena such as
the Stoner criterion for itinerant ferromagnetism or BCS superconductivity, the propensity for FS instabilities
generically depends on the ordering's channel-specific coupling strength
and the density of states (DoS) at
the Fermi level~\cite{Stoner1938,Bardeen1957}. The latter is intuitive from a Hamiltonian spectral
flow perspective: The more spectral weight is pushed away from the
Fermi level the more the ground state can be pushed down in energy.
Van-Hove singularities (vHs) at or close to the Fermi level are particularly useful to elevate instability
scales of electronic order, and are a crucial ingredient to theories
of electronic materials. While the clean limit of vHs in two spatial dimensions
features a logarithmically diverging DoS at specific points
in momentum space, their realization in an imperfect quantum electronic
material still ensures a peak in DoS at the vH points.  

As many electronic kinetic models
naturally feature vH points, it is particularly desirable to
devise ways to tune the vH points of a given band structure 
to the Fermi level in order to facilitate the emergence of vH-driven FS instabilities. Pressure has some tuning impact
on vH points, but also implies side-effects to electronic bands
some of which would reduce the FS instability scales by reducing the
interaction vs. bandwidth ratio~\cite{Consiglio2021}. Doping is another established way of
tuning the vH profile. Unless it is constrained to a few percent of
doping that could be accomplished via electrolytic gating, however,
chemical doping likewise has negative side-effects on FS instability scales such as the
enhancement of disorder~\cite{McChesney2010, Rosenzweig2020, Haberer2013}. 

In this Letter, we propose an alternative principle to perform vH
engineering in a model of interacting electrons on lattices with honeycomb structure. Rather than external pressure or doping, we propose
\textit{compensated metallicity} to deform a Fermi pocket towards the vH points around pristine half filling. We show how the
absence of chiral symmetry due to intra-sublattice hybridization facilitates the emergence of a
circular electron pocket~(ep) at the Brillouin zone~(BZ) center and an enlarged hole pocket~(hp) tuned into proximity to the vH points at~$M$. 
While the scattering vertex contributions of the ep turn out to be of
marginal significance
due to its isotropic character, the vH-proximitized hp triggers ideal
conditions for the unfolding of unconventional
FS instabilities. 
Deriving from ab-inito simulations we substantiate our proposed charge compensation principle by suggesting strained Xene structures \cite{Zhu2015, Molle2017} (\textit{i.e.} silicene, germanene, stanene) as possible material realizations.
\paragraph{Honeycomb Hubbard model} 
To illustrate vH-tuning assisted by compensated
metallicity, we study the $t_1$-$t_2$-$U$ tight-binding Hubbard model on the honeycomb lattice
\begin{equation} \label{eqn:model}
\begin{split}
    \hat H = \Big( - & \,t_1
    \sum_{\langle i,j\rangle \sigma}
    \hat c^\dagger_{i\sigma} \hat c^{\vphantom\dagger}_{j\sigma}
      - t_2
    \sum_{\langle \langle i,j \rangle \rangle \sigma}
    \hat c^\dagger_{i\sigma} \hat c^{\vphantom\dagger}_{j\sigma}  \Big)
    + \text{h.c.} \\
    - & \mu
    \sum_{i \sigma} \hat n_{i\sigma}
    + \, U \sum_{i}
    \hat n_{i\uparrow} \hat n_{i\downarrow} \, , \\
\end{split}
\end{equation}
where $\hat c_{i \sigma}^{(\dagger)}$ denotes electron annihilation (creation)
operators at site $i$ with spin $\sigma$ and $\hat n_{i\sigma}= \hat
c_{i\sigma}^\dagger \hat c_{i\sigma}^{\vphantom\dagger}$,
$U > 0$ is the onsite Coulomb repulsion, and $\mu$ denotes the chemical potential.
The (next) nearest neighbor hopping elements $t_1$ ($t_2$) $>0$ are visualized in~\autoref{fig:bandstructure}(a) as to how they realize (intra-) inter-sublattice~hybridization. 

Eq.~\ref{eqn:model} provides an effective model for graphene monolayers~\cite{Neto2009, Jung2013}. 
Diagonalizing the kinetic part of Eq.~\ref{eqn:model} in the limit $t_2 \ll t_1$, where there is only orbital hybridization between the $p_z$ orbitals of neighboring carbon atoms, a large nested Fermi pocket is realized at van-Hove fillings \mbox{$n = 3/4 , 5/4$}, with a point-like FS hosting Dirac points at half filling $n=1$.
These values are pinned by the chiral symmetry of the bipartite
honeycomb lattice, $\big\{ \hat{H}, \hat C \big\} = 0$, where $\hat{C}
=\hat{P} \cdot \hat{T}$ is defined as the combination of particle-hole
($\hat P$) and time-reversal ($\hat T$) symmetry and can be represented as the
third Pauli matrix acting on the sublattice degree of freedom in
Eq.~\eqref{eqn:model}~\cite{Hatsugai2013}. The chiral symmetry of the energy spectrum is broken by imposing finite intra-sublattice hybridization $t_2$ in~\autoref{fig:bandstructure}(b).

Due to the high quality of graphene samples, where lattice distortions
or vacancies are rare, the clean limit given in Eq.~\eqref{eqn:model} bears
significant resemblance to the experimental setting. A major factor
inhibiting the observation of a FS instability in graphene is the
small DoS around pristine filling. Tuning graphene to a vH level is therefore an alluring path towards enhancing electronic order, and has been extensively discussed in the literature as a
route to promote electronically mediated pairing~\cite{Nandkishore2012, Kiesel2012, Wu2013, Wang2012, Alsharari_2022, Ribeiro2023}.
It is expected to yield a chiral $d$-wave gapped superconductor, which is a topologically ordered quantum
many-body state, \ie, a Chern-Bogoliubov band insulator with Majorana edge modes~\cite{Schaffer2007,Schaffer2012, Can2021, Wolf2022, Crepieux2023}.
The realization of such a state has, however, thus far been
elusive, since homogeneously doping graphene to such high
fillings unavoidably induces a critical amount of disorder to suppress such desired electronic instabilities~\cite{McChesney2010, Rosenzweig2020}.
\paragraph{Van-Hove tuning from compensated metallicity}
As opposed to the single pocket case, a multi-pocket fermiology can enable the transfer of charges between Fermi sheets with electron and hole character, in compliance with the Luttinger theorem~\cite{Luttinger1960a, Luttinger1960b}.
In the last decade, this has been extensively discussed in the case of the iron pnictides~\cite{Chen2008, Chen2008a, Rotter2008}.
In such compounds, an increase of the number of conduction band electrons is compensated for by increasing the occupation of hole states in the valence band. This creates a compensated metal -- an overall charge neutral compound with a large ensemble of
low-energy hole-type and electron-type excitations at pristine filling,
and possibly enhanced nesting effects between eps and hps at and off
the Fermi level~\cite{Chubukov2008, Maiti2010, Chubukov2012}.\\
We find that a key step along the path towards compensated metal
phases in a honecomb lattice model described by
Eq.~\ref{eqn:model} is the assumption of sizable $t_2$, which creates
an asymmetry between the two vHs (\autoref{fig:bandstructure}(b)).
The energy bands are flattened and pushed up in energy at the BZ
boundary, while significant $t_2$ exhibits the opposite effect around the BZ center $\Gamma$, introducing an electron pocket at half-filling for $t_2 > t_1 / 3$. 
By inducing this Lifshitz transition, we allow the
point-like FS of the Dirac point to grow and expand around $K$, eventually approaching a perfectly nested FS coinciding with the lower vHs for $t_2 \sim  0.85 \, t_1$ (\autoref{fig:bandstructure}(c)).
This respects the Luttinger theorem, as in a multi-pocket
scenario of hole-type and charge-type excitations occupied states can be transferred between the pockets while keeping the total electron number constant. Holes in the lower band around $K$ are created by shifting the electrons into the upper band at the BZ center.
We can interpret $t_2$ as a pocket-dependent
chemical potential shift, since adjusting $t_2$ has an effective doping effect on the pocket at
the BZ boundary.
\begin{figure*}[t]
    \centering
    \includegraphics[width = 0.97\textwidth]{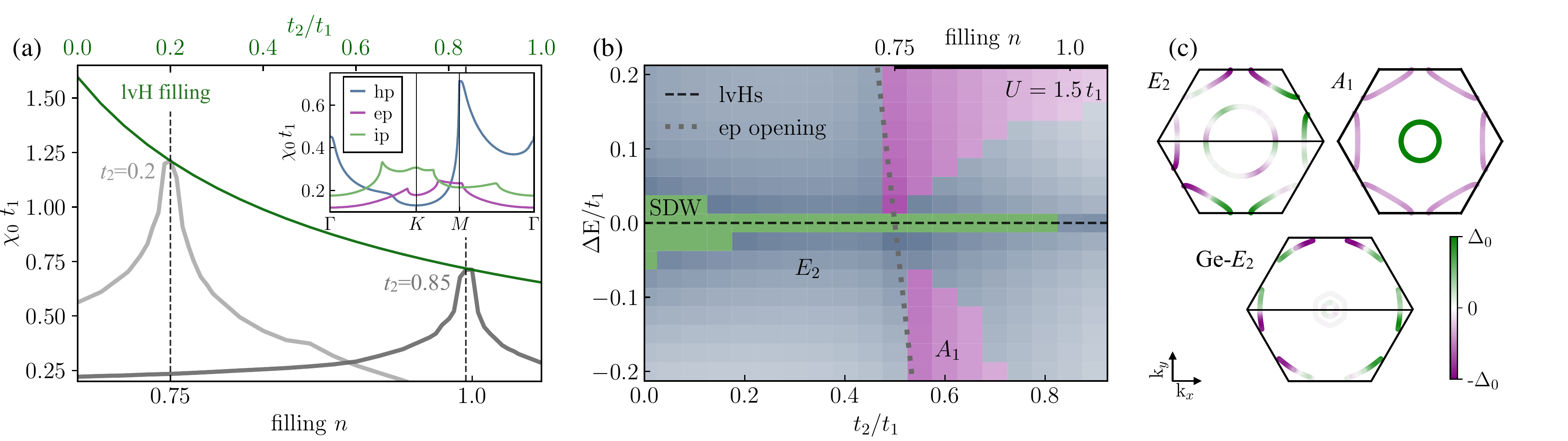}
    \caption{\textit{Many body analysis at lower van-Hove singularity.} (a) Leading eigenvalue of the bare susceptibility $\chi_0$
    (Eq.~\eqref{eqn:bare_susc}).
    Both the dependence on $t_2$ along lower van-Hove (lvH) filling and the doping dependence at fixed long range hopping is depicted.
    The inset shows the full band-resolved fluctuation spectrum including intra-pocket (ep and hp) and inter-pocket (ip) contributions at $t_2=0.85 \, t_1$ and $n=1.0$.
    (b)~Resulting phases for Fermi level detuning $\Delta E$ from lvHs at intermediate interaction scale $U = 1.5\, t_1$.
    The intensity of the color indicates the SC pairing strength, which is proportional to the transition temperature $T_c$.
    (c)~Superconducting gaps of the chiral $d$-wave ($E_2$) and $s_{\pm}$ ($A_1$) phase, respectively.
    Each of the two leading basis functions of the inversion odd $E_2$ irrep is shown in one half of the BZ.
    Almost identical to the leading $E_2$ gap function for Germanene structure tuned to vH filling via compressive strain (Ge-$E_2$, details in SM), the superconducting pairing is exclusively situated on the outer vH pocket.
    All calculations were performed with $( 800 \times 800 )$ integral points and an inverse temperature $\beta = 250 / t_1$ to obtain converged results.
}
\label{fig:fermiology}
\end{figure*}
\paragraph{Electronic response at vH filling}
The combined effect of electronic availability of the vHs and nesting features present on the FS
determine viable electronic fluctuation channels, manifest in the static electronic response function
\begin{equation} \label{eqn:bare_susc}
\begin{split}
    \rchi^0_{o_1o_2o_3o_4}(\mathbf{q}) & = \\
    - \frac{1}{\beta} \sum_n
    \int_\text{BZ}\,\frac{\text{d}\mathbf k}{V_\text{BZ}} &
    G_{o_2o_4}(\mathbf{k}, \omega_n)
    G_{o_3o_1}(\mathbf{k} + \mathbf{q}, \omega_n)
    + \text{h.c.}
\end{split}
\end{equation}
for the screening of the Coulomb interaction.
Here, $G_{o_1o_2}(\mathbf k, \omega_n)$ is the single particle propagator with momentum $\mathbf k$, fermionic Matsubara frequency $\omega_n$ and additional generalized quantum numbers $o_i$, incorporating spin and sublattice degrees of freedom.
The sum over all frequencies and momenta is normalized by the inverse temperature $\beta = (k_B T)^{-1}$ and the BZ volume.

The primary contributions to the susceptibility are made by states
close to the Fermi level, which means that $\rchi^0$ can be split into
three contributions: intra-ep, intra-hp, and ep-hp inter-pocket (ip)
scattering channels. We depict this contribution-resolved
susceptibility for the example case \mbox{$t_2 = 0.85 \, t_1$} in the inset
of~\autoref{fig:fermiology}(a), which is where the vH filling
approaches half filling \mbox{$n=1.0$.}
The emerging pocket around $\Gamma$ is a circular ep, where the
near isotropy does not promote any distinct transfer momenta
$\mathbf q$. This manifests itself as an absence of distinct peaks in
the electronic response~\cite{Duerrnagel2022, Nandkishore2012}. 
Unlike the Fermiology of pnictides, no nesting exists between the two FS sheets, such that inter-pocket scattering processes are marginal.
Hence the dominant fluctuations are exclusively located in the intra-hp
scattering channel, where the vHs tuning provides a large DoS to participate in the electronic ordering at the Fermi level, which is concentrated at the $M$ points of the hexagonal BZ.
Accordingly, fluctuations with wavevector $\mathbf q = M$ contain scattering events with high occupancy of the initial and final states, and thus provide the low-lying fluctuations in the system.

We depict the strength of $M$ point electronic fluctuations $\rchi^0(M)$ at lower van-Hove (lvH) filling as a measure for the emergence of vHs physics in the honeycomb model (\autoref{fig:fermiology}(a)).
The decreasing DoS at lvH filling, as a function of $t_2$ (cf.~\autoref{fig:bandstructure}(d)), directly translates to the fluctuation strength.
However, the fluctuation spectrum appears to be much more susceptible
to a violation of the perfect nesting condition: a slight detuning
from perfect vHs filling leads to a substantial decrease in the
fluctuation strength, as exemplified for two fixed values of $t_2$ in~\autoref{fig:fermiology}(a).
Accordingly perfect nesting plays the primary role in the emergence of
electronic order rather than the decay of DoS associated with increasing $t_2$.
\paragraph{Many-body analysis at the compensated metal crossover.}
We demonstrate compensated metal ordering tendencies when the single
particle states in Eq.~\ref{eqn:model} are subjected to Hubbard interactions.
We model many-body effects within the \textit{random phase
  approximation}
(RPA)~\cite{Takimoto2004,Kubo2007,Graser2009,Altmeyer2016,Duerrnagel2022},
which focuses on collective particle-hole (ph) excitations as a driver of symmetry breaking transitions.
Under this assumption, the full hierarchy of electronic screening processes can be cast into a geometric series that is evaluated analytically.
This dresses the ph susceptibility in Eq.~\ref{eqn:bare_susc} to yield the RPA response functions
\begin{equation}
    \rchi^\text{RPA}_{\rm c/s}(\mathbf{q}) =
    [1 \pm \rchi^0(\mathbf{q}) \, U]^{-1} \rchi^0(\mathbf{q})
\label{eqn:rpa_susc}
\end{equation}
for charge and spin excitations, respectively.
These quantities allow us to probe ph instabilities indicated by a divergent susceptibility channel.
At lower scales, however, the RPA provides ph screening for the effective Cooper pair interaction
\begin{equation}
    \Gamma(\mathbf k, \mathbf{k^\prime}) = U
    + U \rchi^\text{RPA}(\mathbf k - \mathbf{k^\prime}) U
    - U \rchi^\text{RPA}(\mathbf k + \mathbf{k^\prime}) U \ ,
\label{eqn:Veff}
\end{equation}
which can be evaluated within mean field theory at the Fermi level to reveal
the resulting superconducting fluctuations~\cite{Duerrnagel2022}.
As a direct consequence of Eq.~\ref{eqn:rpa_susc} and the absence of
sublattice interference on the honeycomb lattice, the charge response is suppressed by a repulsive interaction $U > 0$, and magnetic fluctuations dominate the low energy physics~\cite{Kiesel2012}.
For $\rchi^\text{RPA}_s$, the higher order screening processes generally increase the imbalance between different momentum channels in the susceptibility.
For sufficient interaction scales, $\rchi^\text{RPA}_s$ diverges in correspondence to a generalized Stoner criterion, resulting in a spin density wave (SDW) transition.   

We presume a generic instability analysis in a small energy window $\Delta E$
around the lower vHs in the absence of chiral symmetry allowing for charge compensation. Considering previous studies of the honeycomb
Hubbard model~\cite{Nandkishore2012, Kiesel2012, Wang2012, Black2014}, that consistently find a SDW directly at vH filling,
as well as the parameter range suitable for RPA studies, we fix the
interaction scale to an intermediate value of $U=1.5\,t_1$ to study
the effect of electronic correlations around the Lifshitz transition
to compensated metallicity. 
The minor dependence on the size of the interaction parameter as well as
cross-channel effects beyond the RPA are analyzed in detail within the more
sophisticated \textit{functional renormalization group} (FRG) scheme in
the SM. The FRG allows for an unbiased treatment of superconducting, charge,
and magnetic transitions~\cite{Platt2013, Beyer2022, Profe2024}, that is
well suited to treat the competition between SDW and $d$-wave SC in
graphene at vH filling~\cite{Kiesel2012, Nandkishore2012}.
We find that the results of RPA and FRG are in good qualitative agreement and
only slightly depend on the values of $U$.
This substantiates the intuitive picture of SC ordering in our model
conveyed by our weak coupling analysis employing the RPA formalism.

The landscape of symmetry broken phases around lower vH in~\autoref{fig:fermiology}(b) is dominated by a superconducting phase which transforms according to the $E_2$ irreducible representation (irrep), displayed in~\autoref{fig:fermiology}(c). The energetically favorable chiral superposition of its two dimensional order parameter gaps out regions of high DoS, \textit{i.e.} areas in the vicinity of the $M$ points~\cite{Black2014}.
Our findings agree with previous numerical studies of the $t_2 = 0$ honeycomb
lattice near vH filling and indicate a chiral $d$-wave
SC~\cite{Kiesel2012,Nandkishore2012,Wu2013}, which remains unaltered in the
compensated metal regime. 
Indicated by its small amplitude the ep barely participates in the pairing process.

While the electronic fluctuation spectrum continues to be dominated by the nesting of the hp, the small Fermi velocity of the ep right above the pocket opening leads to a large DoS at the Fermi level centered around $\Gamma$.
In this region, the available condensation energy for gapping out the central pocket competes with the electronic fluctuation scales of the hp.
A competing $s_\pm$ SC state (cf.~\autoref{fig:fermiology}(c)) arises close to the pocket opening and is eventually superseded by the $E_2$ state at sufficient depth of quadratic band dispersion at $\Gamma$.
Since we are interested in the scenario close to pristine filling
around $t_2 \sim 0.85\, t_1$ this competitive regime can be safely
neglected, and vH-type DoS is identified as the main driver for
symmetry breaking in the investigated region of parameter space.
\paragraph{Strain induced charge compensation in Xene materials}

The most promising route towards charge compensation in Dirac semi-metals, is
buckling: Due to the bipartite unit cell, buckling does not lower the symmetry
of the honeycomb layer and provides a generic way to modify monolayer
vdW materials through substrate engineering \cite{Molle2017, Zhu2015}. The decrease of second nearest neighbour
distance naturally increases the long range intra sublattice hybridization and highlights the
potential of 2D honeycomb layers of the 4th main group like Silicene, Germanene
and Stanene, known as \textit{Xenes}, that naturally exhibit buckling due to a $sp_3$-like hybridisation
process. This buckling tendency can be intensified by utilizing compressive
strain originating from a lattice mismatch between Xene-layer and substrate \cite{Acun2015, Zhang2016, Li2020, DiSante2019, DiSante2019_2}.
To elevate the proposed charge compensation scheme from analytically traceable
tight-binding hamiltonians to real materials, we investigate strained
\textit{Xene}-materials by means of density functional theory simulations
provided in the supplementary material (SM).
The strained Xene structures preserve the dominant FS nesting features of the
hexagonal Fermi pocket.
Consequently, the observed superconducting order parameters closely resemble
the paradigmatic example of the honeycomb Hubbard model at $t_2 \sim 0.85\,
t_1$ discussed earlier (compare Fig.~\ref{fig:fermiology}(c)).
In addition, the multi-orbital nature of the correlated manifold has a twofold
beneficial effect on the charge compensation paradigm:
Firstly, additional single particle bands close to the Fermi energy increase the
capacity of the spherical charge reservoir at constant energy, such that the
required $t_2/t_1$ ratio at vHs is heavily reduced and eventually reached by
small amounts of strain, \textit{e.g.} by a suitable lattice
mismatch with the substrate, on which the monolayer is cleaved (cf. SM).
Secondly, the orbital separation between $p_z$ orbitals on the hexagonal pocket
and $sp_2$ derived contributions on the central pockets further suppresses
inter-pocket scattering events and further purifies the vH dominated physics in
Xene-compounds (cf. SM).



\paragraph{Conclusion and outlook}
For electronic honeycomb lattice models we find strong chiral symmetry
breaking due to next-nearest neighbor hopping $t_2$ as a decisive
parameter to tune the filling of the hole-like van-Hove level. The
compensating electron-like circular FS pocket in the BZ center due to the
Lifshitz transition implied by $t_2$ does not significantly
participate in the symmetry breaking mechanism via many-body effects.
Investigating the ordering tendencies of our honeycomb model around
half filling with a hole pocket at van-Hove level, we obtain the
sought after chiral topological superconducting $d+id$ instability at
pristine filling or for moderate doping achievable through methods
such as electrolytic gating.

Facing graphene with $t_2 / t_1 \sim 0.15$~\cite{Jung2013} as the most
appealing electronic honeycomb layer material, this raises the question of how
to increase this ratio to the proposed high values $\gtrsim 0.8$ required to obtain vH physics at pristine filling.
Beyond our proposed scheme of facilitating substrate engineered 2D Xene structures, orbital-selective inter-layer hybridization with a substrate or other
van-der-Waals (vdW) materials can also promote indirect long range tunneling
via external electronic states~\cite{Diaz2015, Diaz2017, Qu2022}.
Another avenue towards strong intra-sublattice hopping on the honeycomb lattice is
the realization of large Wannier orbitals like the Fidget spinner states
originating from a topological obstruction, that omit the exponential
localization of Wannier states at a single lattice site~\cite{Po2018}.

In this work we discuss chiral symmetry breaking via long ranged hybridization as one mediator inducing compensated metallicity on the honeycomb lattice. However, the presented results rely on the systems fermiology rather than the detailed mechanism of charge compensation.
Phenomenologically, this puts compensated metal tuning on the map
of correlated quantum material design, which might render itself
relevant for future endeavors in stacked multi-layer vdW materials~\cite{Andrei2020,Devakul2021}.


\paragraph{Acknowledgement.}
We thank J. He\ss d\"{o}rfer, and F. Reinert for discussions.
The work is funded by the Deutsche Forschungsgemeinschaft (DFG, German Research
Foundation) through Project-ID 258499086 - SFB 1170, and through the research
unit QUAST, FOR 5249-449872909 (Project 3 and 5), and through the W\"urzburg-Dresden
Cluster of Excellence on Complexity and Topology in Quantum Matter --
\textit{ct.qmat} Project-ID 390858490 - EXC 2147.
M.D. received additional funding from the European Research Council under Grant
No. 771503 (TopMech-Mat).
We acknowledge HPC resources provided by the Erlangen National High Performance
Computing Center (NHR@FAU) of the Friedrich-Alexander-Universit\"at
Erlangen-N\"urnberg (FAU). NHR funding is provided by federal and Bavarian
state authorities. NHR@FAU hardware is partially funded by the DFG – 440719683.
The authors gratefully acknowledge the Gauss Centre for Supercomputing e.V.
(\url{https://www.gauss-centre.eu}) for funding this project by providing computing
time on the GCS Supercomputer SuperMUC-NG at Leibniz Supercomputing Centre
(\url{https://www.lrz.de}).


\let\oldaddcontentsline\addcontentsline
\renewcommand{\addcontentsline}[3]{}
\bibliography{Bibliography}

\begin{thebibliography}{63}%
\makeatletter
\providecommand \@ifxundefined [1]{%
 \@ifx{#1\undefined}
}%
\providecommand \@ifnum [1]{%
 \ifnum #1\expandafter \@firstoftwo
 \else \expandafter \@secondoftwo
 \fi
}%
\providecommand \@ifx [1]{%
 \ifx #1\expandafter \@firstoftwo
 \else \expandafter \@secondoftwo
 \fi
}%
\providecommand \natexlab [1]{#1}%
\providecommand \enquote  [1]{``#1''}%
\providecommand \bibnamefont  [1]{#1}%
\providecommand \bibfnamefont [1]{#1}%
\providecommand \citenamefont [1]{#1}%
\providecommand \href@noop [0]{\@secondoftwo}%
\providecommand \href [0]{\begingroup \@sanitize@url \@href}%
\providecommand \@href[1]{\@@startlink{#1}\@@href}%
\providecommand \@@href[1]{\endgroup#1\@@endlink}%
\providecommand \@sanitize@url [0]{\catcode `\\12\catcode `\$12\catcode
  `\&12\catcode `\#12\catcode `\^12\catcode `\_12\catcode `\%12\relax}%
\providecommand \@@startlink[1]{}%
\providecommand \@@endlink[0]{}%
\providecommand \url  [0]{\begingroup\@sanitize@url \@url }%
\providecommand \@url [1]{\endgroup\@href {#1}{\urlprefix }}%
\providecommand \urlprefix  [0]{URL }%
\providecommand \Eprint [0]{\href }%
\providecommand \doibase [0]{https://doi.org/}%
\providecommand \selectlanguage [0]{\@gobble}%
\providecommand \bibinfo  [0]{\@secondoftwo}%
\providecommand \bibfield  [0]{\@secondoftwo}%
\providecommand \translation [1]{[#1]}%
\providecommand \BibitemOpen [0]{}%
\providecommand \bibitemStop [0]{}%
\providecommand \bibitemNoStop [0]{.\EOS\space}%
\providecommand \EOS [0]{\spacefactor3000\relax}%
\providecommand \BibitemShut  [1]{\csname bibitem#1\endcsname}%
\let\auto@bib@innerbib\@empty
\bibitem [{\citenamefont {Stoner}(1938)}]{Stoner1938}%
  \BibitemOpen
  \bibfield  {author} {\bibinfo {author} {\bibfnamefont {E.~C.}\ \bibnamefont
  {Stoner}},\ }\bibfield  {title} {\bibinfo {title} {Collective electron
  ferromagnetism},\ }\href {https://api.semanticscholar.org/CorpusID:97805670}
  {\bibfield  {journal} {\bibinfo  {journal} {Proceedings of The Royal Society
  A: Mathematical, Physical and Engineering Sciences}\ }\textbf {\bibinfo
  {volume} {165}},\ \bibinfo {pages} {372} (\bibinfo {year}
  {1938})}\BibitemShut {NoStop}%
\bibitem [{\citenamefont {Bardeen}\ \emph {et~al.}(1957)\citenamefont
  {Bardeen}, \citenamefont {Cooper},\ and\ \citenamefont
  {Schrieffer}}]{Bardeen1957}%
  \BibitemOpen
  \bibfield  {author} {\bibinfo {author} {\bibfnamefont {J.}~\bibnamefont
  {Bardeen}}, \bibinfo {author} {\bibfnamefont {L.~N.}\ \bibnamefont
  {Cooper}},\ and\ \bibinfo {author} {\bibfnamefont {J.~R.}\ \bibnamefont
  {Schrieffer}},\ }\bibfield  {title} {\bibinfo {title} {Theory of
  superconductivity},\ }\href {https://doi.org/10.1103/PhysRev.108.1175}
  {\bibfield  {journal} {\bibinfo  {journal} {Phys. Rev.}\ }\textbf {\bibinfo
  {volume} {108}},\ \bibinfo {pages} {1175} (\bibinfo {year}
  {1957})}\BibitemShut {NoStop}%
\bibitem [{\citenamefont {Consiglio}\ \emph {et~al.}(2022)\citenamefont
  {Consiglio}, \citenamefont {Schwemmer}, \citenamefont {Wu}, \citenamefont
  {Hanke}, \citenamefont {Neupert}, \citenamefont {Thomale}, \citenamefont
  {Sangiovanni},\ and\ \citenamefont {Di~Sante}}]{Consiglio2021}%
  \BibitemOpen
  \bibfield  {author} {\bibinfo {author} {\bibfnamefont {A.}~\bibnamefont
  {Consiglio}}, \bibinfo {author} {\bibfnamefont {T.}~\bibnamefont
  {Schwemmer}}, \bibinfo {author} {\bibfnamefont {X.}~\bibnamefont {Wu}},
  \bibinfo {author} {\bibfnamefont {W.}~\bibnamefont {Hanke}}, \bibinfo
  {author} {\bibfnamefont {T.}~\bibnamefont {Neupert}}, \bibinfo {author}
  {\bibfnamefont {R.}~\bibnamefont {Thomale}}, \bibinfo {author} {\bibfnamefont
  {G.}~\bibnamefont {Sangiovanni}},\ and\ \bibinfo {author} {\bibfnamefont
  {D.}~\bibnamefont {Di~Sante}},\ }\bibfield  {title} {\bibinfo {title} {Van
  hove tuning of a${\mathrm{v}}_{3}{\mathrm{sb}}_{5}$ kagome metals under
  pressure and strain},\ }\href {https://doi.org/10.1103/PhysRevB.105.165146}
  {\bibfield  {journal} {\bibinfo  {journal} {Phys. Rev. B}\ }\textbf {\bibinfo
  {volume} {105}},\ \bibinfo {pages} {165146} (\bibinfo {year}
  {2022})}\BibitemShut {NoStop}%
\bibitem [{\citenamefont {McChesney}\ \emph {et~al.}(2010)\citenamefont
  {McChesney}, \citenamefont {Bostwick}, \citenamefont {Ohta}, \citenamefont
  {Seyller}, \citenamefont {Horn}, \citenamefont {Gonz\'alez},\ and\
  \citenamefont {Rotenberg}}]{McChesney2010}%
  \BibitemOpen
  \bibfield  {author} {\bibinfo {author} {\bibfnamefont {J.~L.}\ \bibnamefont
  {McChesney}}, \bibinfo {author} {\bibfnamefont {A.}~\bibnamefont {Bostwick}},
  \bibinfo {author} {\bibfnamefont {T.}~\bibnamefont {Ohta}}, \bibinfo {author}
  {\bibfnamefont {T.}~\bibnamefont {Seyller}}, \bibinfo {author} {\bibfnamefont
  {K.}~\bibnamefont {Horn}}, \bibinfo {author} {\bibfnamefont {J.}~\bibnamefont
  {Gonz\'alez}},\ and\ \bibinfo {author} {\bibfnamefont {E.}~\bibnamefont
  {Rotenberg}},\ }\bibfield  {title} {\bibinfo {title} {Extended van hove
  singularity and superconducting instability in doped graphene},\ }\href
  {https://doi.org/10.1103/PhysRevLett.104.136803} {\bibfield  {journal}
  {\bibinfo  {journal} {Phys. Rev. Lett.}\ }\textbf {\bibinfo {volume} {104}},\
  \bibinfo {pages} {136803} (\bibinfo {year} {2010})}\BibitemShut {NoStop}%
\bibitem [{\citenamefont {Rosenzweig}\ \emph {et~al.}(2020)\citenamefont
  {Rosenzweig}, \citenamefont {Karakachian}, \citenamefont {Marchenko},
  \citenamefont {K\"uster},\ and\ \citenamefont {Starke}}]{Rosenzweig2020}%
  \BibitemOpen
  \bibfield  {author} {\bibinfo {author} {\bibfnamefont {P.}~\bibnamefont
  {Rosenzweig}}, \bibinfo {author} {\bibfnamefont {H.}~\bibnamefont
  {Karakachian}}, \bibinfo {author} {\bibfnamefont {D.}~\bibnamefont
  {Marchenko}}, \bibinfo {author} {\bibfnamefont {K.}~\bibnamefont
  {K\"uster}},\ and\ \bibinfo {author} {\bibfnamefont {U.}~\bibnamefont
  {Starke}},\ }\bibfield  {title} {\bibinfo {title} {Overdoping graphene beyond
  the van hove singularity},\ }\href
  {https://doi.org/10.1103/PhysRevLett.125.176403} {\bibfield  {journal}
  {\bibinfo  {journal} {Phys. Rev. Lett.}\ }\textbf {\bibinfo {volume} {125}},\
  \bibinfo {pages} {176403} (\bibinfo {year} {2020})}\BibitemShut {NoStop}%
\bibitem [{\citenamefont {Haberer}\ \emph {et~al.}(2013)\citenamefont
  {Haberer}, \citenamefont {Petaccia}, \citenamefont {Fedorov}, \citenamefont
  {Praveen}, \citenamefont {Fabris}, \citenamefont {Piccinin}, \citenamefont
  {Vilkov}, \citenamefont {Vyalikh}, \citenamefont {Preobrajenski},
  \citenamefont {Verbitskiy}, \citenamefont {Shiozawa}, \citenamefont {Fink},
  \citenamefont {Knupfer}, \citenamefont {B\"uchner},\ and\ \citenamefont
  {Gr\"uneis}}]{Haberer2013}%
  \BibitemOpen
  \bibfield  {author} {\bibinfo {author} {\bibfnamefont {D.}~\bibnamefont
  {Haberer}}, \bibinfo {author} {\bibfnamefont {L.}~\bibnamefont {Petaccia}},
  \bibinfo {author} {\bibfnamefont {A.~V.}\ \bibnamefont {Fedorov}}, \bibinfo
  {author} {\bibfnamefont {C.~S.}\ \bibnamefont {Praveen}}, \bibinfo {author}
  {\bibfnamefont {S.}~\bibnamefont {Fabris}}, \bibinfo {author} {\bibfnamefont
  {S.}~\bibnamefont {Piccinin}}, \bibinfo {author} {\bibfnamefont
  {O.}~\bibnamefont {Vilkov}}, \bibinfo {author} {\bibfnamefont {D.~V.}\
  \bibnamefont {Vyalikh}}, \bibinfo {author} {\bibfnamefont {A.}~\bibnamefont
  {Preobrajenski}}, \bibinfo {author} {\bibfnamefont {N.~I.}\ \bibnamefont
  {Verbitskiy}}, \bibinfo {author} {\bibfnamefont {H.}~\bibnamefont
  {Shiozawa}}, \bibinfo {author} {\bibfnamefont {J.}~\bibnamefont {Fink}},
  \bibinfo {author} {\bibfnamefont {M.}~\bibnamefont {Knupfer}}, \bibinfo
  {author} {\bibfnamefont {B.}~\bibnamefont {B\"uchner}},\ and\ \bibinfo
  {author} {\bibfnamefont {A.}~\bibnamefont {Gr\"uneis}},\ }\bibfield  {title}
  {\bibinfo {title} {Anisotropic eliashberg function and electron-phonon
  coupling in doped graphene},\ }\href
  {https://doi.org/10.1103/PhysRevB.88.081401} {\bibfield  {journal} {\bibinfo
  {journal} {Phys. Rev. B}\ }\textbf {\bibinfo {volume} {88}},\ \bibinfo
  {pages} {081401} (\bibinfo {year} {2013})}\BibitemShut {NoStop}%
\bibitem [{\citenamefont {Zhu}\ \emph {et~al.}(2015)\citenamefont {Zhu},
  \citenamefont {Chen}, \citenamefont {Xu}, \citenamefont {Gao}, \citenamefont
  {Guan}, \citenamefont {Liu}, \citenamefont {Qian}, \citenamefont {Zhang},\
  and\ \citenamefont {Jia}}]{Zhu2015}%
  \BibitemOpen
  \bibfield  {author} {\bibinfo {author} {\bibfnamefont {F.-f.}\ \bibnamefont
  {Zhu}}, \bibinfo {author} {\bibfnamefont {W.-j.}\ \bibnamefont {Chen}},
  \bibinfo {author} {\bibfnamefont {Y.}~\bibnamefont {Xu}}, \bibinfo {author}
  {\bibfnamefont {C.-l.}\ \bibnamefont {Gao}}, \bibinfo {author} {\bibfnamefont
  {D.-d.}\ \bibnamefont {Guan}}, \bibinfo {author} {\bibfnamefont {C.-h.}\
  \bibnamefont {Liu}}, \bibinfo {author} {\bibfnamefont {D.}~\bibnamefont
  {Qian}}, \bibinfo {author} {\bibfnamefont {S.-C.}\ \bibnamefont {Zhang}},\
  and\ \bibinfo {author} {\bibfnamefont {J.-f.}\ \bibnamefont {Jia}},\
  }\bibfield  {title} {\bibinfo {title} {Epitaxial growth of two-dimensional
  stanene},\ }\href {https://doi.org/10.1038/nmat4384} {\bibfield  {journal}
  {\bibinfo  {journal} {Nature Materials}\ }\textbf {\bibinfo {volume} {14}},\
  \bibinfo {pages} {1020} (\bibinfo {year} {2015})}\BibitemShut {NoStop}%
\bibitem [{\citenamefont {Molle}\ \emph {et~al.}(2017)\citenamefont {Molle},
  \citenamefont {Goldberger}, \citenamefont {Houssa}, \citenamefont {Xu},
  \citenamefont {Zhang},\ and\ \citenamefont {Akinwande}}]{Molle2017}%
  \BibitemOpen
  \bibfield  {author} {\bibinfo {author} {\bibfnamefont {A.}~\bibnamefont
  {Molle}}, \bibinfo {author} {\bibfnamefont {J.}~\bibnamefont {Goldberger}},
  \bibinfo {author} {\bibfnamefont {M.}~\bibnamefont {Houssa}}, \bibinfo
  {author} {\bibfnamefont {Y.}~\bibnamefont {Xu}}, \bibinfo {author}
  {\bibfnamefont {S.-C.}\ \bibnamefont {Zhang}},\ and\ \bibinfo {author}
  {\bibfnamefont {D.}~\bibnamefont {Akinwande}},\ }\bibfield  {title} {\bibinfo
  {title} {Buckled two-dimensional xene sheets},\ }\href
  {https://doi.org/10.1038/nmat4802} {\bibfield  {journal} {\bibinfo  {journal}
  {Nature Materials}\ }\textbf {\bibinfo {volume} {16}},\ \bibinfo {pages}
  {163} (\bibinfo {year} {2017})}\BibitemShut {NoStop}%
\bibitem [{\citenamefont {Castro~Neto}\ \emph {et~al.}(2009)\citenamefont
  {Castro~Neto}, \citenamefont {Guinea}, \citenamefont {Peres}, \citenamefont
  {Novoselov},\ and\ \citenamefont {Geim}}]{Neto2009}%
  \BibitemOpen
  \bibfield  {author} {\bibinfo {author} {\bibfnamefont {A.~H.}\ \bibnamefont
  {Castro~Neto}}, \bibinfo {author} {\bibfnamefont {F.}~\bibnamefont {Guinea}},
  \bibinfo {author} {\bibfnamefont {N.~M.~R.}\ \bibnamefont {Peres}}, \bibinfo
  {author} {\bibfnamefont {K.~S.}\ \bibnamefont {Novoselov}},\ and\ \bibinfo
  {author} {\bibfnamefont {A.~K.}\ \bibnamefont {Geim}},\ }\bibfield  {title}
  {\bibinfo {title} {The electronic properties of graphene},\ }\href
  {https://doi.org/10.1103/RevModPhys.81.109} {\bibfield  {journal} {\bibinfo
  {journal} {Rev. Mod. Phys.}\ }\textbf {\bibinfo {volume} {81}},\ \bibinfo
  {pages} {109} (\bibinfo {year} {2009})}\BibitemShut {NoStop}%
\bibitem [{\citenamefont {Jung}\ and\ \citenamefont
  {MacDonald}(2013)}]{Jung2013}%
  \BibitemOpen
  \bibfield  {author} {\bibinfo {author} {\bibfnamefont {J.}~\bibnamefont
  {Jung}}\ and\ \bibinfo {author} {\bibfnamefont {A.~H.}\ \bibnamefont
  {MacDonald}},\ }\bibfield  {title} {\bibinfo {title} {Tight-binding model for
  graphene $\ensuremath{\pi}$-bands from maximally localized wannier
  functions},\ }\href {https://doi.org/10.1103/PhysRevB.87.195450} {\bibfield
  {journal} {\bibinfo  {journal} {Phys. Rev. B}\ }\textbf {\bibinfo {volume}
  {87}},\ \bibinfo {pages} {195450} (\bibinfo {year} {2013})}\BibitemShut
  {NoStop}%
\bibitem [{\citenamefont {Hatsugai}\ \emph {et~al.}(2013)\citenamefont
  {Hatsugai}, \citenamefont {Morimoto}, \citenamefont {Kawarabayashi},
  \citenamefont {Hamamoto},\ and\ \citenamefont {Aoki}}]{Hatsugai2013}%
  \BibitemOpen
  \bibfield  {author} {\bibinfo {author} {\bibfnamefont {Y.}~\bibnamefont
  {Hatsugai}}, \bibinfo {author} {\bibfnamefont {T.}~\bibnamefont {Morimoto}},
  \bibinfo {author} {\bibfnamefont {T.}~\bibnamefont {Kawarabayashi}}, \bibinfo
  {author} {\bibfnamefont {Y.}~\bibnamefont {Hamamoto}},\ and\ \bibinfo
  {author} {\bibfnamefont {H.}~\bibnamefont {Aoki}},\ }\bibfield  {title}
  {\bibinfo {title} {Chiral symmetry and its manifestation in optical responses
  in graphene: interaction and multilayers},\ }\href
  {https://doi.org/10.1088/1367-2630/15/3/035023} {\bibfield  {journal}
  {\bibinfo  {journal} {New Journal of Physics}\ }\textbf {\bibinfo {volume}
  {15}},\ \bibinfo {pages} {035023} (\bibinfo {year} {2013})}\BibitemShut
  {NoStop}%
\bibitem [{\citenamefont {Nandkishore}\ \emph {et~al.}(2012)\citenamefont
  {Nandkishore}, \citenamefont {Levitov},\ and\ \citenamefont
  {Chubukov}}]{Nandkishore2012}%
  \BibitemOpen
  \bibfield  {author} {\bibinfo {author} {\bibfnamefont {R.}~\bibnamefont
  {Nandkishore}}, \bibinfo {author} {\bibfnamefont {L.~S.}\ \bibnamefont
  {Levitov}},\ and\ \bibinfo {author} {\bibfnamefont {A.~V.}\ \bibnamefont
  {Chubukov}},\ }\bibfield  {title} {\bibinfo {title} {Chiral superconductivity
  from repulsive interactions in doped graphene},\ }\href
  {https://doi.org/10.1038/nphys2208} {\bibfield  {journal} {\bibinfo
  {journal} {Nature Physics}\ }\textbf {\bibinfo {volume} {8}},\ \bibinfo
  {pages} {158} (\bibinfo {year} {2012})}\BibitemShut {NoStop}%
\bibitem [{\citenamefont {Kiesel}\ \emph {et~al.}(2012)\citenamefont {Kiesel},
  \citenamefont {Platt}, \citenamefont {Hanke}, \citenamefont {Abanin},\ and\
  \citenamefont {Thomale}}]{Kiesel2012}%
  \BibitemOpen
  \bibfield  {author} {\bibinfo {author} {\bibfnamefont {M.~L.}\ \bibnamefont
  {Kiesel}}, \bibinfo {author} {\bibfnamefont {C.}~\bibnamefont {Platt}},
  \bibinfo {author} {\bibfnamefont {W.}~\bibnamefont {Hanke}}, \bibinfo
  {author} {\bibfnamefont {D.~A.}\ \bibnamefont {Abanin}},\ and\ \bibinfo
  {author} {\bibfnamefont {R.}~\bibnamefont {Thomale}},\ }\bibfield  {title}
  {\bibinfo {title} {Competing many-body instabilities and unconventional
  superconductivity in graphene},\ }\href
  {https://doi.org/10.1103/PhysRevB.86.020507} {\bibfield  {journal} {\bibinfo
  {journal} {Phys. Rev. B}\ }\textbf {\bibinfo {volume} {86}},\ \bibinfo
  {pages} {020507} (\bibinfo {year} {2012})}\BibitemShut {NoStop}%
\bibitem [{\citenamefont {Wu}\ \emph {et~al.}(2013)\citenamefont {Wu},
  \citenamefont {Scherer}, \citenamefont {Honerkamp},\ and\ \citenamefont
  {Le~Hur}}]{Wu2013}%
  \BibitemOpen
  \bibfield  {author} {\bibinfo {author} {\bibfnamefont {W.}~\bibnamefont
  {Wu}}, \bibinfo {author} {\bibfnamefont {M.~M.}\ \bibnamefont {Scherer}},
  \bibinfo {author} {\bibfnamefont {C.}~\bibnamefont {Honerkamp}},\ and\
  \bibinfo {author} {\bibfnamefont {K.}~\bibnamefont {Le~Hur}},\ }\bibfield
  {title} {\bibinfo {title} {Correlated {{Dirac}} particles and
  superconductivity on the honeycomb lattice},\ }\href
  {https://doi.org/10.1103/PhysRevB.87.094521} {\bibfield  {journal} {\bibinfo
  {journal} {Phys. Rev. B}\ }\textbf {\bibinfo {volume} {87}},\ \bibinfo
  {pages} {094521} (\bibinfo {year} {2013})}\BibitemShut {NoStop}%
\bibitem [{\citenamefont {Wang}\ \emph {et~al.}(2012)\citenamefont {Wang},
  \citenamefont {Xiang}, \citenamefont {Wang}, \citenamefont {Wang},
  \citenamefont {Yang},\ and\ \citenamefont {Lee}}]{Wang2012}%
  \BibitemOpen
  \bibfield  {author} {\bibinfo {author} {\bibfnamefont {W.-S.}\ \bibnamefont
  {Wang}}, \bibinfo {author} {\bibfnamefont {Y.-Y.}\ \bibnamefont {Xiang}},
  \bibinfo {author} {\bibfnamefont {Q.-H.}\ \bibnamefont {Wang}}, \bibinfo
  {author} {\bibfnamefont {F.}~\bibnamefont {Wang}}, \bibinfo {author}
  {\bibfnamefont {F.}~\bibnamefont {Yang}},\ and\ \bibinfo {author}
  {\bibfnamefont {D.-H.}\ \bibnamefont {Lee}},\ }\bibfield  {title} {\bibinfo
  {title} {Functional renormalization group and variational monte carlo studies
  of the electronic instabilities in graphene near $\frac{1}{4}$ doping},\
  }\href {https://doi.org/10.1103/PhysRevB.85.035414} {\bibfield  {journal}
  {\bibinfo  {journal} {Phys. Rev. B}\ }\textbf {\bibinfo {volume} {85}},\
  \bibinfo {pages} {035414} (\bibinfo {year} {2012})}\BibitemShut {NoStop}%
\bibitem [{\citenamefont {Alsharari}\ and\ \citenamefont
  {Ulloa}(2022)}]{Alsharari_2022}%
  \BibitemOpen
  \bibfield  {author} {\bibinfo {author} {\bibfnamefont {A.~M.}\ \bibnamefont
  {Alsharari}}\ and\ \bibinfo {author} {\bibfnamefont {S.~E.}\ \bibnamefont
  {Ulloa}},\ }\bibfield  {title} {\bibinfo {title} {Inducing chiral
  superconductivity on honeycomb lattice systems},\ }\href
  {https://doi.org/10.1088/1361-648X/ac5a03} {\bibfield  {journal} {\bibinfo
  {journal} {Journal of Physics: Condensed Matter}\ }\textbf {\bibinfo {volume}
  {34}},\ \bibinfo {pages} {205403} (\bibinfo {year} {2022})}\BibitemShut
  {NoStop}%
\bibitem [{\citenamefont {Ribeiro}\ \emph {et~al.}(2023)\citenamefont
  {Ribeiro}, \citenamefont {Raposo},\ and\ \citenamefont
  {Coutinho-Filho}}]{Ribeiro2023}%
  \BibitemOpen
  \bibfield  {author} {\bibinfo {author} {\bibfnamefont {F.~G.}\ \bibnamefont
  {Ribeiro}}, \bibinfo {author} {\bibfnamefont {E.~P.}\ \bibnamefont
  {Raposo}},\ and\ \bibinfo {author} {\bibfnamefont {M.~D.}\ \bibnamefont
  {Coutinho-Filho}},\ }\bibfield  {title} {\bibinfo {title} {Competing chiral
  $d$-wave superconductivity and magnetic phases in the strong-coupling hubbard
  model on the honeycomb lattice},\ }\href
  {https://doi.org/10.1103/PhysRevB.107.064510} {\bibfield  {journal} {\bibinfo
   {journal} {Phys. Rev. B}\ }\textbf {\bibinfo {volume} {107}},\ \bibinfo
  {pages} {064510} (\bibinfo {year} {2023})}\BibitemShut {NoStop}%
\bibitem [{\citenamefont {Black-Schaffer}\ and\ \citenamefont
  {Doniach}(2007)}]{Schaffer2007}%
  \BibitemOpen
  \bibfield  {author} {\bibinfo {author} {\bibfnamefont {A.~M.}\ \bibnamefont
  {Black-Schaffer}}\ and\ \bibinfo {author} {\bibfnamefont {S.}~\bibnamefont
  {Doniach}},\ }\bibfield  {title} {\bibinfo {title} {Resonating valence bonds
  and mean-field $d$-wave superconductivity in graphite},\ }\href
  {https://doi.org/10.1103/PhysRevB.75.134512} {\bibfield  {journal} {\bibinfo
  {journal} {Phys. Rev. B}\ }\textbf {\bibinfo {volume} {75}},\ \bibinfo
  {pages} {134512} (\bibinfo {year} {2007})}\BibitemShut {NoStop}%
\bibitem [{\citenamefont {Black-Schaffer}(2012)}]{Schaffer2012}%
  \BibitemOpen
  \bibfield  {author} {\bibinfo {author} {\bibfnamefont {A.~M.}\ \bibnamefont
  {Black-Schaffer}},\ }\bibfield  {title} {\bibinfo {title} {Edge properties
  and majorana fermions in the proposed chiral $d$-wave superconducting state
  of doped graphene},\ }\href {https://doi.org/10.1103/PhysRevLett.109.197001}
  {\bibfield  {journal} {\bibinfo  {journal} {Phys. Rev. Lett.}\ }\textbf
  {\bibinfo {volume} {109}},\ \bibinfo {pages} {197001} (\bibinfo {year}
  {2012})}\BibitemShut {NoStop}%
\bibitem [{\citenamefont {Can}\ \emph {et~al.}(2021)\citenamefont {Can},
  \citenamefont {Tummuru}, \citenamefont {Day}, \citenamefont {Elfimov},
  \citenamefont {Damascelli},\ and\ \citenamefont {Franz}}]{Can2021}%
  \BibitemOpen
  \bibfield  {author} {\bibinfo {author} {\bibfnamefont {O.}~\bibnamefont
  {Can}}, \bibinfo {author} {\bibfnamefont {T.}~\bibnamefont {Tummuru}},
  \bibinfo {author} {\bibfnamefont {R.~P.}\ \bibnamefont {Day}}, \bibinfo
  {author} {\bibfnamefont {I.}~\bibnamefont {Elfimov}}, \bibinfo {author}
  {\bibfnamefont {A.}~\bibnamefont {Damascelli}},\ and\ \bibinfo {author}
  {\bibfnamefont {M.}~\bibnamefont {Franz}},\ }\bibfield  {title} {\bibinfo
  {title} {High-temperature topological superconductivity in twisted
  double-layer copper oxides},\ }\href
  {https://doi.org/10.1038/s41567-020-01142-7} {\bibfield  {journal} {\bibinfo
  {journal} {Nature Physics}\ }\textbf {\bibinfo {volume} {17}},\ \bibinfo
  {pages} {519} (\bibinfo {year} {2021})}\BibitemShut {NoStop}%
\bibitem [{\citenamefont {Wolf}\ \emph
  {et~al.}(2022{\natexlab{a}})\citenamefont {Wolf}, \citenamefont {Gardener},
  \citenamefont {Hur},\ and\ \citenamefont {Rachel}}]{Wolf2022}%
  \BibitemOpen
  \bibfield  {author} {\bibinfo {author} {\bibfnamefont {S.}~\bibnamefont
  {Wolf}}, \bibinfo {author} {\bibfnamefont {T.}~\bibnamefont {Gardener}},
  \bibinfo {author} {\bibfnamefont {K.~L.}\ \bibnamefont {Hur}},\ and\ \bibinfo
  {author} {\bibfnamefont {S.}~\bibnamefont {Rachel}},\ }\bibfield  {title}
  {\bibinfo {title} {Topological superconductivity on the honeycomb lattice:
  {{Effect}} of normal state topology},\ }\href
  {https://doi.org/10.1103/PhysRevB.105.L100505} {\bibfield  {journal}
  {\bibinfo  {journal} {Phys. Rev. B}\ }\textbf {\bibinfo {volume} {105}},\
  \bibinfo {pages} {L100505} (\bibinfo {year}
  {2022}{\natexlab{a}})}\BibitemShut {NoStop}%
\bibitem [{\citenamefont {Cr\'epieux}\ \emph {et~al.}(2023)\citenamefont
  {Cr\'epieux}, \citenamefont {Pangburn}, \citenamefont {Haurie}, \citenamefont
  {Awoga}, \citenamefont {Black-Schaffer}, \citenamefont {Sedlmayr},
  \citenamefont {P\'epin},\ and\ \citenamefont {Bena}}]{Crepieux2023}%
  \BibitemOpen
  \bibfield  {author} {\bibinfo {author} {\bibfnamefont {A.}~\bibnamefont
  {Cr\'epieux}}, \bibinfo {author} {\bibfnamefont {E.}~\bibnamefont
  {Pangburn}}, \bibinfo {author} {\bibfnamefont {L.}~\bibnamefont {Haurie}},
  \bibinfo {author} {\bibfnamefont {O.~A.}\ \bibnamefont {Awoga}}, \bibinfo
  {author} {\bibfnamefont {A.~M.}\ \bibnamefont {Black-Schaffer}}, \bibinfo
  {author} {\bibfnamefont {N.}~\bibnamefont {Sedlmayr}}, \bibinfo {author}
  {\bibfnamefont {C.}~\bibnamefont {P\'epin}},\ and\ \bibinfo {author}
  {\bibfnamefont {C.}~\bibnamefont {Bena}},\ }\bibfield  {title} {\bibinfo
  {title} {Superconductivity in monolayer and few-layer graphene. ii.
  topological edge states and chern numbers},\ }\href
  {https://doi.org/10.1103/PhysRevB.108.134515} {\bibfield  {journal} {\bibinfo
   {journal} {Phys. Rev. B}\ }\textbf {\bibinfo {volume} {108}},\ \bibinfo
  {pages} {134515} (\bibinfo {year} {2023})}\BibitemShut {NoStop}%
\bibitem [{\citenamefont {Luttinger}\ and\ \citenamefont
  {Ward}(1960)}]{Luttinger1960a}%
  \BibitemOpen
  \bibfield  {author} {\bibinfo {author} {\bibfnamefont {J.~M.}\ \bibnamefont
  {Luttinger}}\ and\ \bibinfo {author} {\bibfnamefont {J.~C.}\ \bibnamefont
  {Ward}},\ }\bibfield  {title} {\bibinfo {title} {Ground-state energy of a
  many-fermion system. ii},\ }\href {https://doi.org/10.1103/PhysRev.118.1417}
  {\bibfield  {journal} {\bibinfo  {journal} {Phys. Rev.}\ }\textbf {\bibinfo
  {volume} {118}},\ \bibinfo {pages} {1417} (\bibinfo {year}
  {1960})}\BibitemShut {NoStop}%
\bibitem [{\citenamefont {Luttinger}(1960)}]{Luttinger1960b}%
  \BibitemOpen
  \bibfield  {author} {\bibinfo {author} {\bibfnamefont {J.~M.}\ \bibnamefont
  {Luttinger}},\ }\bibfield  {title} {\bibinfo {title} {Fermi surface and some
  simple equilibrium properties of a system of interacting fermions},\ }\href
  {https://doi.org/10.1103/PhysRev.119.1153} {\bibfield  {journal} {\bibinfo
  {journal} {Phys. Rev.}\ }\textbf {\bibinfo {volume} {119}},\ \bibinfo {pages}
  {1153} (\bibinfo {year} {1960})}\BibitemShut {NoStop}%
\bibitem [{\citenamefont {Chen}\ \emph
  {et~al.}(2008{\natexlab{a}})\citenamefont {Chen}, \citenamefont {Li},
  \citenamefont {Wu}, \citenamefont {Li}, \citenamefont {Hu}, \citenamefont
  {Dong}, \citenamefont {Zheng}, \citenamefont {Luo},\ and\ \citenamefont
  {Wang}}]{Chen2008}%
  \BibitemOpen
  \bibfield  {author} {\bibinfo {author} {\bibfnamefont {G.~F.}\ \bibnamefont
  {Chen}}, \bibinfo {author} {\bibfnamefont {Z.}~\bibnamefont {Li}}, \bibinfo
  {author} {\bibfnamefont {D.}~\bibnamefont {Wu}}, \bibinfo {author}
  {\bibfnamefont {G.}~\bibnamefont {Li}}, \bibinfo {author} {\bibfnamefont
  {W.~Z.}\ \bibnamefont {Hu}}, \bibinfo {author} {\bibfnamefont
  {J.}~\bibnamefont {Dong}}, \bibinfo {author} {\bibfnamefont {P.}~\bibnamefont
  {Zheng}}, \bibinfo {author} {\bibfnamefont {J.~L.}\ \bibnamefont {Luo}},\
  and\ \bibinfo {author} {\bibfnamefont {N.~L.}\ \bibnamefont {Wang}},\
  }\bibfield  {title} {\bibinfo {title} {Superconductivity at 41 k and its
  competition with spin-density-wave instability in layered
  ${\mathrm{ceo}}_{1\ensuremath{-}x}{\mathrm{f}}_{x}\mathrm{FeAs}$},\ }\href
  {https://doi.org/10.1103/PhysRevLett.100.247002} {\bibfield  {journal}
  {\bibinfo  {journal} {Phys. Rev. Lett.}\ }\textbf {\bibinfo {volume} {100}},\
  \bibinfo {pages} {247002} (\bibinfo {year} {2008}{\natexlab{a}})}\BibitemShut
  {NoStop}%
\bibitem [{\citenamefont {Chen}\ \emph
  {et~al.}(2008{\natexlab{b}})\citenamefont {Chen}, \citenamefont {Wu},
  \citenamefont {Wu}, \citenamefont {Liu}, \citenamefont {Chen},\ and\
  \citenamefont {Fang}}]{Chen2008a}%
  \BibitemOpen
  \bibfield  {author} {\bibinfo {author} {\bibfnamefont {X.~H.}\ \bibnamefont
  {Chen}}, \bibinfo {author} {\bibfnamefont {T.}~\bibnamefont {Wu}}, \bibinfo
  {author} {\bibfnamefont {G.}~\bibnamefont {Wu}}, \bibinfo {author}
  {\bibfnamefont {R.~H.}\ \bibnamefont {Liu}}, \bibinfo {author} {\bibfnamefont
  {H.}~\bibnamefont {Chen}},\ and\ \bibinfo {author} {\bibfnamefont {D.~F.}\
  \bibnamefont {Fang}},\ }\bibfield  {title} {\bibinfo {title}
  {Superconductivity at 43{\thinspace}k in smfeaso1-xf x},\ }\href
  {https://doi.org/10.1038/nature07045} {\bibfield  {journal} {\bibinfo
  {journal} {Nature}\ }\textbf {\bibinfo {volume} {453}},\ \bibinfo {pages}
  {761} (\bibinfo {year} {2008}{\natexlab{b}})}\BibitemShut {NoStop}%
\bibitem [{\citenamefont {Rotter}\ \emph {et~al.}(2008)\citenamefont {Rotter},
  \citenamefont {Tegel},\ and\ \citenamefont {Johrendt}}]{Rotter2008}%
  \BibitemOpen
  \bibfield  {author} {\bibinfo {author} {\bibfnamefont {M.}~\bibnamefont
  {Rotter}}, \bibinfo {author} {\bibfnamefont {M.}~\bibnamefont {Tegel}},\ and\
  \bibinfo {author} {\bibfnamefont {D.}~\bibnamefont {Johrendt}},\ }\bibfield
  {title} {\bibinfo {title} {Superconductivity at 38 k in the iron arsenide
  $({\mathrm{ba}}_{1\ensuremath{-}x}{\mathrm{k}}_{x}){\mathrm{fe}}_{2}{\mathrm{as}}_{2}$},\
  }\href {https://doi.org/10.1103/PhysRevLett.101.107006} {\bibfield  {journal}
  {\bibinfo  {journal} {Phys. Rev. Lett.}\ }\textbf {\bibinfo {volume} {101}},\
  \bibinfo {pages} {107006} (\bibinfo {year} {2008})}\BibitemShut {NoStop}%
\bibitem [{\citenamefont {Chubukov}\ \emph {et~al.}(2008)\citenamefont
  {Chubukov}, \citenamefont {Efremov},\ and\ \citenamefont
  {Eremin}}]{Chubukov2008}%
  \BibitemOpen
  \bibfield  {author} {\bibinfo {author} {\bibfnamefont {A.~V.}\ \bibnamefont
  {Chubukov}}, \bibinfo {author} {\bibfnamefont {D.~V.}\ \bibnamefont
  {Efremov}},\ and\ \bibinfo {author} {\bibfnamefont {I.}~\bibnamefont
  {Eremin}},\ }\bibfield  {title} {\bibinfo {title} {Magnetism,
  superconductivity, and pairing symmetry in iron-based superconductors},\
  }\href {https://doi.org/10.1103/PhysRevB.78.134512} {\bibfield  {journal}
  {\bibinfo  {journal} {Phys. Rev. B}\ }\textbf {\bibinfo {volume} {78}},\
  \bibinfo {pages} {134512} (\bibinfo {year} {2008})}\BibitemShut {NoStop}%
\bibitem [{\citenamefont {Maiti}\ and\ \citenamefont
  {Chubukov}(2010)}]{Maiti2010}%
  \BibitemOpen
  \bibfield  {author} {\bibinfo {author} {\bibfnamefont {S.}~\bibnamefont
  {Maiti}}\ and\ \bibinfo {author} {\bibfnamefont {A.~V.}\ \bibnamefont
  {Chubukov}},\ }\bibfield  {title} {\bibinfo {title} {Renormalization group
  flow, competing phases, and the structure of superconducting gap in multiband
  models of iron-based superconductors},\ }\href
  {https://doi.org/10.1103/PhysRevB.82.214515} {\bibfield  {journal} {\bibinfo
  {journal} {Phys. Rev. B}\ }\textbf {\bibinfo {volume} {82}},\ \bibinfo
  {pages} {214515} (\bibinfo {year} {2010})}\BibitemShut {NoStop}%
\bibitem [{\citenamefont {Chubukov}(2012)}]{Chubukov2012}%
  \BibitemOpen
  \bibfield  {author} {\bibinfo {author} {\bibfnamefont {A.}~\bibnamefont
  {Chubukov}},\ }\bibfield  {title} {\bibinfo {title} {Pairing mechanism in
  fe-based superconductors},\ }\href
  {https://doi.org/10.1146/annurev-conmatphys-020911-125055} {\bibfield
  {journal} {\bibinfo  {journal} {Annual Review of Condensed Matter Physics}\
  }\textbf {\bibinfo {volume} {3}},\ \bibinfo {pages} {57} (\bibinfo {year}
  {2012})}\BibitemShut {NoStop}%
\bibitem [{\citenamefont {Dürrnagel}\ \emph {et~al.}(2022)\citenamefont
  {Dürrnagel}, \citenamefont {Beyer}, \citenamefont {Thomale},\ and\
  \citenamefont {Schwemmer}}]{Duerrnagel2022}%
  \BibitemOpen
  \bibfield  {author} {\bibinfo {author} {\bibfnamefont {M.}~\bibnamefont
  {Dürrnagel}}, \bibinfo {author} {\bibfnamefont {J.}~\bibnamefont {Beyer}},
  \bibinfo {author} {\bibfnamefont {R.}~\bibnamefont {Thomale}},\ and\ \bibinfo
  {author} {\bibfnamefont {T.}~\bibnamefont {Schwemmer}},\ }\bibfield  {title}
  {\bibinfo {title} {Unconventional superconductivity from weak coupling - a
  unified perspective on formalism and numerical implementation},\ }\href
  {https://doi.org/10.1140/epjb/s10051-022-00371-4} {\bibfield  {journal}
  {\bibinfo  {journal} {The European Physical Journal B}\ }\textbf {\bibinfo
  {volume} {95}},\ \bibinfo {pages} {112} (\bibinfo {year} {2022})}\BibitemShut
  {NoStop}%
\bibitem [{\citenamefont {Takimoto}\ \emph {et~al.}(2004)\citenamefont
  {Takimoto}, \citenamefont {Hotta},\ and\ \citenamefont
  {Ueda}}]{Takimoto2004}%
  \BibitemOpen
  \bibfield  {author} {\bibinfo {author} {\bibfnamefont {T.}~\bibnamefont
  {Takimoto}}, \bibinfo {author} {\bibfnamefont {T.}~\bibnamefont {Hotta}},\
  and\ \bibinfo {author} {\bibfnamefont {K.}~\bibnamefont {Ueda}},\ }\bibfield
  {title} {\bibinfo {title} {Strong-coupling theory of superconductivity in a
  degenerate hubbard model},\ }\href
  {https://doi.org/10.1103/PhysRevB.69.104504} {\bibfield  {journal} {\bibinfo
  {journal} {Phys. Rev. B}\ }\textbf {\bibinfo {volume} {69}},\ \bibinfo
  {pages} {104504} (\bibinfo {year} {2004})}\BibitemShut {NoStop}%
\bibitem [{\citenamefont {Kubo}(2007)}]{Kubo2007}%
  \BibitemOpen
  \bibfield  {author} {\bibinfo {author} {\bibfnamefont {K.}~\bibnamefont
  {Kubo}},\ }\bibfield  {title} {\bibinfo {title} {Pairing symmetry in a
  two-orbital hubbard model on a square lattice},\ }\href
  {https://doi.org/10.1103/PhysRevB.75.224509} {\bibfield  {journal} {\bibinfo
  {journal} {Phys. Rev. B}\ }\textbf {\bibinfo {volume} {75}},\ \bibinfo
  {pages} {224509} (\bibinfo {year} {2007})}\BibitemShut {NoStop}%
\bibitem [{\citenamefont {Graser}\ \emph {et~al.}(2009)\citenamefont {Graser},
  \citenamefont {Maier}, \citenamefont {Hirschfeld},\ and\ \citenamefont
  {Scalapino}}]{Graser2009}%
  \BibitemOpen
  \bibfield  {author} {\bibinfo {author} {\bibfnamefont {S.}~\bibnamefont
  {Graser}}, \bibinfo {author} {\bibfnamefont {T.~A.}\ \bibnamefont {Maier}},
  \bibinfo {author} {\bibfnamefont {P.~J.}\ \bibnamefont {Hirschfeld}},\ and\
  \bibinfo {author} {\bibfnamefont {D.~J.}\ \bibnamefont {Scalapino}},\
  }\bibfield  {title} {\bibinfo {title} {Near-degeneracy of several pairing
  channels in multiorbital models for the fe pnictides},\ }\href
  {https://doi.org/10.1088/1367-2630/11/2/025016} {\bibfield  {journal}
  {\bibinfo  {journal} {New Journal of Physics}\ }\textbf {\bibinfo {volume}
  {11}},\ \bibinfo {pages} {025016} (\bibinfo {year} {2009})}\BibitemShut
  {NoStop}%
\bibitem [{\citenamefont {Altmeyer}\ \emph {et~al.}(2016)\citenamefont
  {Altmeyer}, \citenamefont {Guterding}, \citenamefont {Hirschfeld},
  \citenamefont {Maier}, \citenamefont {Valent\'{\i}},\ and\ \citenamefont
  {Scalapino}}]{Altmeyer2016}%
  \BibitemOpen
  \bibfield  {author} {\bibinfo {author} {\bibfnamefont {M.}~\bibnamefont
  {Altmeyer}}, \bibinfo {author} {\bibfnamefont {D.}~\bibnamefont {Guterding}},
  \bibinfo {author} {\bibfnamefont {P.~J.}\ \bibnamefont {Hirschfeld}},
  \bibinfo {author} {\bibfnamefont {T.~A.}\ \bibnamefont {Maier}}, \bibinfo
  {author} {\bibfnamefont {R.}~\bibnamefont {Valent\'{\i}}},\ and\ \bibinfo
  {author} {\bibfnamefont {D.~J.}\ \bibnamefont {Scalapino}},\ }\bibfield
  {title} {\bibinfo {title} {Role of vertex corrections in the matrix
  formulation of the random phase approximation for the multiorbital hubbard
  model},\ }\href {https://doi.org/10.1103/PhysRevB.94.214515} {\bibfield
  {journal} {\bibinfo  {journal} {Phys. Rev. B}\ }\textbf {\bibinfo {volume}
  {94}},\ \bibinfo {pages} {214515} (\bibinfo {year} {2016})}\BibitemShut
  {NoStop}%
\bibitem [{\citenamefont {Black-Schaffer}\ and\ \citenamefont
  {Honerkamp}(2014)}]{Black2014}%
  \BibitemOpen
  \bibfield  {author} {\bibinfo {author} {\bibfnamefont {A.~M.}\ \bibnamefont
  {Black-Schaffer}}\ and\ \bibinfo {author} {\bibfnamefont {C.}~\bibnamefont
  {Honerkamp}},\ }\bibfield  {title} {\bibinfo {title} {Chiral d-wave
  superconductivity in doped graphene},\ }\href
  {https://doi.org/10.1088/0953-8984/26/42/423201} {\bibfield  {journal}
  {\bibinfo  {journal} {Journal of Physics: Condensed Matter}\ }\textbf
  {\bibinfo {volume} {26}},\ \bibinfo {pages} {423201} (\bibinfo {year}
  {2014})}\BibitemShut {NoStop}%
\bibitem [{\citenamefont {C.~Platt}\ and\ \citenamefont
  {Thomale}(2013)}]{Platt2013}%
  \BibitemOpen
  \bibfield  {author} {\bibinfo {author} {\bibfnamefont {W.~H.}\ \bibnamefont
  {C.~Platt}}\ and\ \bibinfo {author} {\bibfnamefont {R.}~\bibnamefont
  {Thomale}},\ }\bibfield  {title} {\bibinfo {title} {Functional
  renormalization group for multi-orbital fermi surface instabilities},\ }\href
  {https://doi.org/10.1080/00018732.2013.862020} {\bibfield  {journal}
  {\bibinfo  {journal} {Advances in Physics}\ }\textbf {\bibinfo {volume}
  {62}},\ \bibinfo {pages} {453} (\bibinfo {year} {2013})}\BibitemShut
  {NoStop}%
\bibitem [{\citenamefont {Beyer}\ \emph {et~al.}(2022)\citenamefont {Beyer},
  \citenamefont {Hauck},\ and\ \citenamefont {Klebl}}]{Beyer2022}%
  \BibitemOpen
  \bibfield  {author} {\bibinfo {author} {\bibfnamefont {J.}~\bibnamefont
  {Beyer}}, \bibinfo {author} {\bibfnamefont {J.~B.}\ \bibnamefont {Hauck}},\
  and\ \bibinfo {author} {\bibfnamefont {L.}~\bibnamefont {Klebl}},\ }\bibfield
   {title} {\bibinfo {title} {Reference results for the momentum space
  functional renormalization group},\ }\href
  {https://doi.org/10.1140/epjb/s10051-022-00323-y} {\bibfield  {journal}
  {\bibinfo  {journal} {The European Physical Journal B}\ }\textbf {\bibinfo
  {volume} {95}},\ \bibinfo {pages} {65} (\bibinfo {year} {2022})}\BibitemShut
  {NoStop}%
\bibitem [{\citenamefont {Profe}\ \emph {et~al.}(2024)\citenamefont {Profe},
  \citenamefont {Rhodes}, \citenamefont {D\"urrnagel}, \citenamefont {Bisset},
  \citenamefont {Marques}, \citenamefont {Chi}, \citenamefont {Schwemmer},
  \citenamefont {Thomale}, \citenamefont {Kennes}, \citenamefont {Hooley},\
  and\ \citenamefont {Wahl}}]{Profe2024}%
  \BibitemOpen
  \bibfield  {author} {\bibinfo {author} {\bibfnamefont {J.~B.}\ \bibnamefont
  {Profe}}, \bibinfo {author} {\bibfnamefont {L.~C.}\ \bibnamefont {Rhodes}},
  \bibinfo {author} {\bibfnamefont {M.}~\bibnamefont {D\"urrnagel}}, \bibinfo
  {author} {\bibfnamefont {R.}~\bibnamefont {Bisset}}, \bibinfo {author}
  {\bibfnamefont {C.~A.}\ \bibnamefont {Marques}}, \bibinfo {author}
  {\bibfnamefont {S.}~\bibnamefont {Chi}}, \bibinfo {author} {\bibfnamefont
  {T.}~\bibnamefont {Schwemmer}}, \bibinfo {author} {\bibfnamefont
  {R.}~\bibnamefont {Thomale}}, \bibinfo {author} {\bibfnamefont {D.~M.}\
  \bibnamefont {Kennes}}, \bibinfo {author} {\bibfnamefont {C.~A.}\
  \bibnamefont {Hooley}},\ and\ \bibinfo {author} {\bibfnamefont
  {P.}~\bibnamefont {Wahl}},\ }\bibfield  {title} {\bibinfo {title} {Magic
  angle of ${\mathrm{sr}}_{2}{\mathrm{ruo}}_{4}$: Optimizing correlation-driven
  superconductivity},\ }\href
  {https://doi.org/10.1103/PhysRevResearch.6.043057} {\bibfield  {journal}
  {\bibinfo  {journal} {Phys. Rev. Res.}\ }\textbf {\bibinfo {volume} {6}},\
  \bibinfo {pages} {043057} (\bibinfo {year} {2024})}\BibitemShut {NoStop}%
\bibitem [{\citenamefont {Acun}\ \emph {et~al.}(2015)\citenamefont {Acun},
  \citenamefont {Zhang}, \citenamefont {Bampoulis}, \citenamefont {Farmanbar},
  \citenamefont {Houselt}, \citenamefont {Rudenko}, \citenamefont
  {Lingenfelder}, \citenamefont {Brocks}, \citenamefont {Poelsema},
  \citenamefont {Katsnelson},\ and\ \citenamefont {Zandvliet}}]{Acun2015}%
  \BibitemOpen
  \bibfield  {author} {\bibinfo {author} {\bibfnamefont {A.}~\bibnamefont
  {Acun}}, \bibinfo {author} {\bibfnamefont {L.}~\bibnamefont {Zhang}},
  \bibinfo {author} {\bibfnamefont {P.}~\bibnamefont {Bampoulis}}, \bibinfo
  {author} {\bibfnamefont {M.}~\bibnamefont {Farmanbar}}, \bibinfo {author}
  {\bibfnamefont {A.}~\bibnamefont {Houselt}}, \bibinfo {author} {\bibfnamefont
  {A.}~\bibnamefont {Rudenko}}, \bibinfo {author} {\bibfnamefont
  {M.}~\bibnamefont {Lingenfelder}}, \bibinfo {author} {\bibfnamefont
  {G.}~\bibnamefont {Brocks}}, \bibinfo {author} {\bibfnamefont
  {B.}~\bibnamefont {Poelsema}}, \bibinfo {author} {\bibfnamefont
  {M.}~\bibnamefont {Katsnelson}},\ and\ \bibinfo {author} {\bibfnamefont
  {H.}~\bibnamefont {Zandvliet}},\ }\bibfield  {title} {\bibinfo {title}
  {Germanene: The germanium analogue of graphene},\ }\href
  {https://doi.org/10.1088/0953-8984/27/44/443002} {\bibfield  {journal}
  {\bibinfo  {journal} {Journal of physics. Condensed matter : an Institute of
  Physics journal}\ }\textbf {\bibinfo {volume} {27}},\ \bibinfo {pages}
  {443002} (\bibinfo {year} {2015})}\BibitemShut {NoStop}%
\bibitem [{\citenamefont {Zhang}\ \emph {et~al.}(2016)\citenamefont {Zhang},
  \citenamefont {Bampoulis}, \citenamefont {Rudenko}, \citenamefont {Yao},
  \citenamefont {van Houselt}, \citenamefont {Poelsema}, \citenamefont
  {Katsnelson},\ and\ \citenamefont {Zandvliet}}]{Zhang2016}%
  \BibitemOpen
  \bibfield  {author} {\bibinfo {author} {\bibfnamefont {L.}~\bibnamefont
  {Zhang}}, \bibinfo {author} {\bibfnamefont {P.}~\bibnamefont {Bampoulis}},
  \bibinfo {author} {\bibfnamefont {A.~N.}\ \bibnamefont {Rudenko}}, \bibinfo
  {author} {\bibfnamefont {Q.}~\bibnamefont {Yao}}, \bibinfo {author}
  {\bibfnamefont {A.}~\bibnamefont {van Houselt}}, \bibinfo {author}
  {\bibfnamefont {B.}~\bibnamefont {Poelsema}}, \bibinfo {author}
  {\bibfnamefont {M.~I.}\ \bibnamefont {Katsnelson}},\ and\ \bibinfo {author}
  {\bibfnamefont {H.~J.~W.}\ \bibnamefont {Zandvliet}},\ }\bibfield  {title}
  {\bibinfo {title} {Structural and electronic properties of germanene on
  ${\mathrm{mos}}_{2}$},\ }\href
  {https://doi.org/10.1103/PhysRevLett.116.256804} {\bibfield  {journal}
  {\bibinfo  {journal} {Phys. Rev. Lett.}\ }\textbf {\bibinfo {volume} {116}},\
  \bibinfo {pages} {256804} (\bibinfo {year} {2016})}\BibitemShut {NoStop}%
\bibitem [{\citenamefont {Li}\ \emph {et~al.}(2020)\citenamefont {Li},
  \citenamefont {Lei}, \citenamefont {Wang}, \citenamefont {Wu}, \citenamefont
  {Zhao}, \citenamefont {Zhao}, \citenamefont {Guo}, \citenamefont {Qian},\
  and\ \citenamefont {Ibrahim}}]{Li2020}%
  \BibitemOpen
  \bibfield  {author} {\bibinfo {author} {\bibfnamefont {J.}~\bibnamefont
  {Li}}, \bibinfo {author} {\bibfnamefont {T.}~\bibnamefont {Lei}}, \bibinfo
  {author} {\bibfnamefont {J.}~\bibnamefont {Wang}}, \bibinfo {author}
  {\bibfnamefont {R.}~\bibnamefont {Wu}}, \bibinfo {author} {\bibfnamefont
  {J.}~\bibnamefont {Zhao}}, \bibinfo {author} {\bibfnamefont {L.}~\bibnamefont
  {Zhao}}, \bibinfo {author} {\bibfnamefont {Y.}~\bibnamefont {Guo}}, \bibinfo
  {author} {\bibfnamefont {H.}~\bibnamefont {Qian}},\ and\ \bibinfo {author}
  {\bibfnamefont {K.}~\bibnamefont {Ibrahim}},\ }\bibfield  {title} {\bibinfo
  {title} {Anisotropic electronic structure and interfacial chemical reaction
  of stanene/bi2te3},\ }\href {https://doi.org/10.1021/acs.jpcc.0c00139}
  {\bibfield  {journal} {\bibinfo  {journal} {The Journal of Physical Chemistry
  C}\ }\textbf {\bibinfo {volume} {124}},\ \bibinfo {pages} {4917} (\bibinfo
  {year} {2020})}\BibitemShut {NoStop}%
\bibitem [{\citenamefont {Di~Sante}\ \emph
  {et~al.}(2019{\natexlab{a}})\citenamefont {Di~Sante}, \citenamefont {Wu},
  \citenamefont {Fink}, \citenamefont {Hanke},\ and\ \citenamefont
  {Thomale}}]{DiSante2019}%
  \BibitemOpen
  \bibfield  {author} {\bibinfo {author} {\bibfnamefont {D.}~\bibnamefont
  {Di~Sante}}, \bibinfo {author} {\bibfnamefont {X.}~\bibnamefont {Wu}},
  \bibinfo {author} {\bibfnamefont {M.}~\bibnamefont {Fink}}, \bibinfo {author}
  {\bibfnamefont {W.}~\bibnamefont {Hanke}},\ and\ \bibinfo {author}
  {\bibfnamefont {R.}~\bibnamefont {Thomale}},\ }\bibfield  {title} {\bibinfo
  {title} {Triplet superconductivity in the dirac semimetal germanene on a
  substrate},\ }\href {https://doi.org/10.1103/PhysRevB.99.201106} {\bibfield
  {journal} {\bibinfo  {journal} {Phys. Rev. B}\ }\textbf {\bibinfo {volume}
  {99}},\ \bibinfo {pages} {201106} (\bibinfo {year}
  {2019}{\natexlab{a}})}\BibitemShut {NoStop}%
\bibitem [{\citenamefont {Di~Sante}\ \emph
  {et~al.}(2019{\natexlab{b}})\citenamefont {Di~Sante}, \citenamefont {Eck},
  \citenamefont {Bauernfeind}, \citenamefont {Will}, \citenamefont {Thomale},
  \citenamefont {Sch\"afer}, \citenamefont {Claessen},\ and\ \citenamefont
  {Sangiovanni}}]{DiSante2019_2}%
  \BibitemOpen
  \bibfield  {author} {\bibinfo {author} {\bibfnamefont {D.}~\bibnamefont
  {Di~Sante}}, \bibinfo {author} {\bibfnamefont {P.}~\bibnamefont {Eck}},
  \bibinfo {author} {\bibfnamefont {M.}~\bibnamefont {Bauernfeind}}, \bibinfo
  {author} {\bibfnamefont {M.}~\bibnamefont {Will}}, \bibinfo {author}
  {\bibfnamefont {R.}~\bibnamefont {Thomale}}, \bibinfo {author} {\bibfnamefont
  {J.}~\bibnamefont {Sch\"afer}}, \bibinfo {author} {\bibfnamefont
  {R.}~\bibnamefont {Claessen}},\ and\ \bibinfo {author} {\bibfnamefont
  {G.}~\bibnamefont {Sangiovanni}},\ }\bibfield  {title} {\bibinfo {title}
  {Towards topological quasifreestanding stanene via substrate engineering},\
  }\href {https://doi.org/10.1103/PhysRevB.99.035145} {\bibfield  {journal}
  {\bibinfo  {journal} {Phys. Rev. B}\ }\textbf {\bibinfo {volume} {99}},\
  \bibinfo {pages} {035145} (\bibinfo {year} {2019}{\natexlab{b}})}\BibitemShut
  {NoStop}%
\bibitem [{\citenamefont {Coy~Diaz}\ \emph {et~al.}(2015)\citenamefont
  {Coy~Diaz}, \citenamefont {Avila}, \citenamefont {Chen}, \citenamefont
  {Addou}, \citenamefont {Asensio},\ and\ \citenamefont {Batzill}}]{Diaz2015}%
  \BibitemOpen
  \bibfield  {author} {\bibinfo {author} {\bibfnamefont {H.}~\bibnamefont
  {Coy~Diaz}}, \bibinfo {author} {\bibfnamefont {J.}~\bibnamefont {Avila}},
  \bibinfo {author} {\bibfnamefont {C.}~\bibnamefont {Chen}}, \bibinfo {author}
  {\bibfnamefont {R.}~\bibnamefont {Addou}}, \bibinfo {author} {\bibfnamefont
  {M.}~\bibnamefont {Asensio}},\ and\ \bibinfo {author} {\bibfnamefont
  {M.}~\bibnamefont {Batzill}},\ }\bibfield  {title} {\bibinfo {title} {Direct
  observation of interlayer hybridization and dirac relativistic carriers in
  graphene/mos 2 van der waals heterostructures},\ }\href
  {https://doi.org/10.1021/nl504167y} {\bibfield  {journal} {\bibinfo
  {journal} {Nano letters}\ }\textbf {\bibinfo {volume} {15}} (\bibinfo {year}
  {2015})}\BibitemShut {NoStop}%
\bibitem [{\citenamefont {Diaz}\ \emph {et~al.}(2017)\citenamefont {Diaz},
  \citenamefont {Ma}, \citenamefont {Kolekar}, \citenamefont {Avila},
  \citenamefont {Chen}, \citenamefont {Asensio},\ and\ \citenamefont
  {Batzill}}]{Diaz2017}%
  \BibitemOpen
  \bibfield  {author} {\bibinfo {author} {\bibfnamefont {H.~C.}\ \bibnamefont
  {Diaz}}, \bibinfo {author} {\bibfnamefont {Y.}~\bibnamefont {Ma}}, \bibinfo
  {author} {\bibfnamefont {S.}~\bibnamefont {Kolekar}}, \bibinfo {author}
  {\bibfnamefont {J.}~\bibnamefont {Avila}}, \bibinfo {author} {\bibfnamefont
  {C.}~\bibnamefont {Chen}}, \bibinfo {author} {\bibfnamefont {M.~C.}\
  \bibnamefont {Asensio}},\ and\ \bibinfo {author} {\bibfnamefont
  {M.}~\bibnamefont {Batzill}},\ }\bibfield  {title} {\bibinfo {title}
  {Substrate dependent electronic structure variations of van der waals
  heterostructures of mose2 or mose2(1$\-$x)te2x grown by van der waals
  epitaxy},\ }\href {https://doi.org/10.1088/2053-1583/aa6e6a} {\bibfield
  {journal} {\bibinfo  {journal} {2D Materials}\ }\textbf {\bibinfo {volume}
  {4}},\ \bibinfo {pages} {025094} (\bibinfo {year} {2017})}\BibitemShut
  {NoStop}%
\bibitem [{\citenamefont {Qu}\ \emph {et~al.}(2022)\citenamefont {Qu},
  \citenamefont {Xu}, \citenamefont {Cao}, \citenamefont {Wang}, \citenamefont
  {Wang}, \citenamefont {Shi},\ and\ \citenamefont {Xu}}]{Qu2022}%
  \BibitemOpen
  \bibfield  {author} {\bibinfo {author} {\bibfnamefont {Y.}~\bibnamefont
  {Qu}}, \bibinfo {author} {\bibfnamefont {Y.}~\bibnamefont {Xu}}, \bibinfo
  {author} {\bibfnamefont {B.}~\bibnamefont {Cao}}, \bibinfo {author}
  {\bibfnamefont {Y.}~\bibnamefont {Wang}}, \bibinfo {author} {\bibfnamefont
  {J.}~\bibnamefont {Wang}}, \bibinfo {author} {\bibfnamefont {L.}~\bibnamefont
  {Shi}},\ and\ \bibinfo {author} {\bibfnamefont {K.}~\bibnamefont {Xu}},\
  }\bibfield  {title} {\bibinfo {title} {Long-range orbital hybridization in
  remote epitaxy: The nucleation mechanism of gan on different substrates via
  single-layer graphene},\ }\href {https://doi.org/10.1021/acsami.1c18926}
  {\bibfield  {journal} {\bibinfo  {journal} {ACS Applied Materials \&
  Interfaces}\ }\textbf {\bibinfo {volume} {14}} (\bibinfo {year}
  {2022})}\BibitemShut {NoStop}%
\bibitem [{\citenamefont {Po}\ \emph {et~al.}(2018)\citenamefont {Po},
  \citenamefont {Zou}, \citenamefont {Vishwanath},\ and\ \citenamefont
  {Senthil}}]{Po2018}%
  \BibitemOpen
  \bibfield  {author} {\bibinfo {author} {\bibfnamefont {H.~C.}\ \bibnamefont
  {Po}}, \bibinfo {author} {\bibfnamefont {L.}~\bibnamefont {Zou}}, \bibinfo
  {author} {\bibfnamefont {A.}~\bibnamefont {Vishwanath}},\ and\ \bibinfo
  {author} {\bibfnamefont {T.}~\bibnamefont {Senthil}},\ }\bibfield  {title}
  {\bibinfo {title} {Origin of mott insulating behavior and superconductivity
  in twisted bilayer graphene},\ }\href
  {https://doi.org/10.1103/PhysRevX.8.031089} {\bibfield  {journal} {\bibinfo
  {journal} {Phys. Rev. X}\ }\textbf {\bibinfo {volume} {8}},\ \bibinfo {pages}
  {031089} (\bibinfo {year} {2018})}\BibitemShut {NoStop}%
\bibitem [{\citenamefont {Andrei}\ and\ \citenamefont
  {MacDonald}(2020)}]{Andrei2020}%
  \BibitemOpen
  \bibfield  {author} {\bibinfo {author} {\bibfnamefont {E.~Y.}\ \bibnamefont
  {Andrei}}\ and\ \bibinfo {author} {\bibfnamefont {A.~H.}\ \bibnamefont
  {MacDonald}},\ }\bibfield  {title} {\bibinfo {title} {Graphene bilayers with
  a twist},\ }\href {https://doi.org/10.1038/s41563-020-00840-0} {\bibfield
  {journal} {\bibinfo  {journal} {Nature Materials}\ }\textbf {\bibinfo
  {volume} {19}},\ \bibinfo {pages} {1265} (\bibinfo {year}
  {2020})}\BibitemShut {NoStop}%
\bibitem [{\citenamefont {Devakul}\ \emph {et~al.}(2021)\citenamefont
  {Devakul}, \citenamefont {Cr{\'e}pel}, \citenamefont {Zhang},\ and\
  \citenamefont {Fu}}]{Devakul2021}%
  \BibitemOpen
  \bibfield  {author} {\bibinfo {author} {\bibfnamefont {T.}~\bibnamefont
  {Devakul}}, \bibinfo {author} {\bibfnamefont {V.}~\bibnamefont {Cr{\'e}pel}},
  \bibinfo {author} {\bibfnamefont {Y.}~\bibnamefont {Zhang}},\ and\ \bibinfo
  {author} {\bibfnamefont {L.}~\bibnamefont {Fu}},\ }\bibfield  {title}
  {\bibinfo {title} {Magic in twisted transition metal dichalcogenide
  bilayers},\ }\href {https://doi.org/10.1038/s41467-021-27042-9} {\bibfield
  {journal} {\bibinfo  {journal} {Nature Communications}\ }\textbf {\bibinfo
  {volume} {12}},\ \bibinfo {pages} {6730} (\bibinfo {year}
  {2021})}\BibitemShut {NoStop}%
\bibitem [{\citenamefont {de~la Pe\~na}\ \emph {et~al.}(2017)\citenamefont
  {de~la Pe\~na}, \citenamefont {Lichtenstein}, \citenamefont {Honerkamp},\
  and\ \citenamefont {Scherer}}]{Pena2017}%
  \BibitemOpen
  \bibfield  {author} {\bibinfo {author} {\bibfnamefont {D.~S.}\ \bibnamefont
  {de~la Pe\~na}}, \bibinfo {author} {\bibfnamefont {J.}~\bibnamefont
  {Lichtenstein}}, \bibinfo {author} {\bibfnamefont {C.}~\bibnamefont
  {Honerkamp}},\ and\ \bibinfo {author} {\bibfnamefont {M.~M.}\ \bibnamefont
  {Scherer}},\ }\bibfield  {title} {\bibinfo {title} {Antiferromagnetism and
  competing charge instabilities of electrons in strained graphene from coulomb
  interactions},\ }\href {https://doi.org/10.1103/PhysRevB.96.205155}
  {\bibfield  {journal} {\bibinfo  {journal} {Phys. Rev. B}\ }\textbf {\bibinfo
  {volume} {96}},\ \bibinfo {pages} {205155} (\bibinfo {year}
  {2017})}\BibitemShut {NoStop}%
\bibitem [{\citenamefont {Meyer}\ \emph {et~al.}(2007)\citenamefont {Meyer},
  \citenamefont {Geim}, \citenamefont {Katsnelson}, \citenamefont {Novoselov},
  \citenamefont {Booth},\ and\ \citenamefont {Roth}}]{Meyer2007}%
  \BibitemOpen
  \bibfield  {author} {\bibinfo {author} {\bibfnamefont {J.~C.}\ \bibnamefont
  {Meyer}}, \bibinfo {author} {\bibfnamefont {A.~K.}\ \bibnamefont {Geim}},
  \bibinfo {author} {\bibfnamefont {M.~I.}\ \bibnamefont {Katsnelson}},
  \bibinfo {author} {\bibfnamefont {K.~S.}\ \bibnamefont {Novoselov}}, \bibinfo
  {author} {\bibfnamefont {T.~J.}\ \bibnamefont {Booth}},\ and\ \bibinfo
  {author} {\bibfnamefont {S.}~\bibnamefont {Roth}},\ }\bibfield  {title}
  {\bibinfo {title} {The structure of suspended graphene sheets},\ }\href
  {https://doi.org/10.1038/nature05545} {\bibfield  {journal} {\bibinfo
  {journal} {Nature}\ }\textbf {\bibinfo {volume} {446}},\ \bibinfo {pages}
  {60–63} (\bibinfo {year} {2007})}\BibitemShut {NoStop}%
\bibitem [{\citenamefont {Wang}\ \emph {et~al.}(2015)\citenamefont {Wang},
  \citenamefont {Chen}, \citenamefont {Wang}, \citenamefont {Cui},\ and\
  \citenamefont {Zhang}}]{Wang2015}%
  \BibitemOpen
  \bibfield  {author} {\bibinfo {author} {\bibfnamefont {D.}~\bibnamefont
  {Wang}}, \bibinfo {author} {\bibfnamefont {L.}~\bibnamefont {Chen}}, \bibinfo
  {author} {\bibfnamefont {X.}~\bibnamefont {Wang}}, \bibinfo {author}
  {\bibfnamefont {G.}~\bibnamefont {Cui}},\ and\ \bibinfo {author}
  {\bibfnamefont {P.}~\bibnamefont {Zhang}},\ }\bibfield  {title} {\bibinfo
  {title} {The effect of substrate and external strain on electronic structures
  of stanene film},\ }\href {https://doi.org/10.1039/c5cp04322k} {\bibfield
  {journal} {\bibinfo  {journal} {Physical chemistry chemical physics : PCCP}\
  }\textbf {\bibinfo {volume} {17}} (\bibinfo {year} {2015})}\BibitemShut
  {NoStop}%
\bibitem [{\citenamefont {Gou}\ \emph {et~al.}(2017)\citenamefont {Gou},
  \citenamefont {Kong}, \citenamefont {Li}, \citenamefont {Zhong},
  \citenamefont {Li}, \citenamefont {Cheng}, \citenamefont {Chen},\ and\
  \citenamefont {Wu}}]{Gou2017}%
  \BibitemOpen
  \bibfield  {author} {\bibinfo {author} {\bibfnamefont {J.}~\bibnamefont
  {Gou}}, \bibinfo {author} {\bibfnamefont {L.}~\bibnamefont {Kong}}, \bibinfo
  {author} {\bibfnamefont {H.}~\bibnamefont {Li}}, \bibinfo {author}
  {\bibfnamefont {Q.}~\bibnamefont {Zhong}}, \bibinfo {author} {\bibfnamefont
  {W.}~\bibnamefont {Li}}, \bibinfo {author} {\bibfnamefont {P.}~\bibnamefont
  {Cheng}}, \bibinfo {author} {\bibfnamefont {L.}~\bibnamefont {Chen}},\ and\
  \bibinfo {author} {\bibfnamefont {K.}~\bibnamefont {Wu}},\ }\bibfield
  {title} {\bibinfo {title} {Strain-induced band engineering in monolayer
  stanene on sb(111)},\ }\href
  {https://doi.org/10.1103/PhysRevMaterials.1.054004} {\bibfield  {journal}
  {\bibinfo  {journal} {Phys. Rev. Mater.}\ }\textbf {\bibinfo {volume} {1}},\
  \bibinfo {pages} {054004} (\bibinfo {year} {2017})}\BibitemShut {NoStop}%
\bibitem [{\citenamefont {Gmitra}\ \emph {et~al.}(2009)\citenamefont {Gmitra},
  \citenamefont {Konschuh}, \citenamefont {Ertler}, \citenamefont
  {Ambrosch-Draxl},\ and\ \citenamefont {Fabian}}]{gmitra2009band}%
  \BibitemOpen
  \bibfield  {author} {\bibinfo {author} {\bibfnamefont {M.}~\bibnamefont
  {Gmitra}}, \bibinfo {author} {\bibfnamefont {S.}~\bibnamefont {Konschuh}},
  \bibinfo {author} {\bibfnamefont {C.}~\bibnamefont {Ertler}}, \bibinfo
  {author} {\bibfnamefont {C.}~\bibnamefont {Ambrosch-Draxl}},\ and\ \bibinfo
  {author} {\bibfnamefont {J.}~\bibnamefont {Fabian}},\ }\bibfield  {title}
  {\bibinfo {title} {Band-structure topologies of graphene: Spin-orbit coupling
  effects from first principles},\ }\href@noop {} {\bibfield  {journal}
  {\bibinfo  {journal} {Phys. Rev. B}\ }\textbf {\bibinfo {volume} {80}},\
  \bibinfo {pages} {235431} (\bibinfo {year} {2009})}\BibitemShut {NoStop}%
\bibitem [{\citenamefont {Bunney}\ \emph {et~al.}(2024)\citenamefont {Bunney},
  \citenamefont {Beyer}, \citenamefont {Thomale}, \citenamefont {Honerkamp},\
  and\ \citenamefont {Rachel}}]{bunney2024chern}%
  \BibitemOpen
  \bibfield  {author} {\bibinfo {author} {\bibfnamefont {M.}~\bibnamefont
  {Bunney}}, \bibinfo {author} {\bibfnamefont {J.}~\bibnamefont {Beyer}},
  \bibinfo {author} {\bibfnamefont {R.}~\bibnamefont {Thomale}}, \bibinfo
  {author} {\bibfnamefont {C.}~\bibnamefont {Honerkamp}},\ and\ \bibinfo
  {author} {\bibfnamefont {S.}~\bibnamefont {Rachel}},\ }\bibfield  {title}
  {\bibinfo {title} {Chern number landscape of spin-orbit coupled chiral
  superconductors},\ }\href@noop {} {\bibfield  {journal} {\bibinfo  {journal}
  {arXiv preprint arXiv:2405.03757}\ } (\bibinfo {year} {2024})}\BibitemShut
  {NoStop}%
\bibitem [{\citenamefont {Wolf}\ \emph
  {et~al.}(2022{\natexlab{b}})\citenamefont {Wolf}, \citenamefont {Gardener},
  \citenamefont {Le~Hur},\ and\ \citenamefont {Rachel}}]{wolf2022topological}%
  \BibitemOpen
  \bibfield  {author} {\bibinfo {author} {\bibfnamefont {S.}~\bibnamefont
  {Wolf}}, \bibinfo {author} {\bibfnamefont {T.}~\bibnamefont {Gardener}},
  \bibinfo {author} {\bibfnamefont {K.}~\bibnamefont {Le~Hur}},\ and\ \bibinfo
  {author} {\bibfnamefont {S.}~\bibnamefont {Rachel}},\ }\bibfield  {title}
  {\bibinfo {title} {Topological superconductivity on the honeycomb lattice:
  Effect of normal state topology},\ }\href@noop {} {\bibfield  {journal}
  {\bibinfo  {journal} {Phys. Rev. B}\ }\textbf {\bibinfo {volume} {105}},\
  \bibinfo {pages} {L100505} (\bibinfo {year}
  {2022}{\natexlab{b}})}\BibitemShut {NoStop}%
\bibitem [{\citenamefont {Kresse}\ and\ \citenamefont
  {Furthm\"uller}(1996)}]{VASP}%
  \BibitemOpen
  \bibfield  {author} {\bibinfo {author} {\bibfnamefont {G.}~\bibnamefont
  {Kresse}}\ and\ \bibinfo {author} {\bibfnamefont {J.}~\bibnamefont
  {Furthm\"uller}},\ }\bibfield  {title} {\bibinfo {title} {Efficient iterative
  schemes for ab initio total-energy calculations using a plane-wave basis
  set},\ }\href {https://doi.org/10.1103/PhysRevB.54.11169} {\bibfield
  {journal} {\bibinfo  {journal} {Phys. Rev. B}\ }\textbf {\bibinfo {volume}
  {54}},\ \bibinfo {pages} {11169} (\bibinfo {year} {1996})}\BibitemShut
  {NoStop}%
\bibitem [{\citenamefont {Kresse}\ and\ \citenamefont {Joubert}(1999)}]{PAW1}%
  \BibitemOpen
  \bibfield  {author} {\bibinfo {author} {\bibfnamefont {G.}~\bibnamefont
  {Kresse}}\ and\ \bibinfo {author} {\bibfnamefont {D.}~\bibnamefont
  {Joubert}},\ }\bibfield  {title} {\bibinfo {title} {From ultrasoft
  pseudopotentials to the projector augmented-wave method},\ }\href
  {https://doi.org/10.1103/PhysRevB.59.1758} {\bibfield  {journal} {\bibinfo
  {journal} {Phys. Rev. B}\ }\textbf {\bibinfo {volume} {59}},\ \bibinfo
  {pages} {1758} (\bibinfo {year} {1999})}\BibitemShut {NoStop}%
\bibitem [{\citenamefont {Bl\"ochl}(1994)}]{PAW2}%
  \BibitemOpen
  \bibfield  {author} {\bibinfo {author} {\bibfnamefont {P.~E.}\ \bibnamefont
  {Bl\"ochl}},\ }\bibfield  {title} {\bibinfo {title} {Projector augmented-wave
  method},\ }\href {https://doi.org/10.1103/PhysRevB.50.17953} {\bibfield
  {journal} {\bibinfo  {journal} {Phys. Rev. B}\ }\textbf {\bibinfo {volume}
  {50}},\ \bibinfo {pages} {17953} (\bibinfo {year} {1994})}\BibitemShut
  {NoStop}%
\bibitem [{\citenamefont {Perdew}\ \emph {et~al.}(1996)\citenamefont {Perdew},
  \citenamefont {Burke},\ and\ \citenamefont {Ernzerhof}}]{PBE}%
  \BibitemOpen
  \bibfield  {author} {\bibinfo {author} {\bibfnamefont {J.~P.}\ \bibnamefont
  {Perdew}}, \bibinfo {author} {\bibfnamefont {K.}~\bibnamefont {Burke}},\ and\
  \bibinfo {author} {\bibfnamefont {M.}~\bibnamefont {Ernzerhof}},\ }\bibfield
  {title} {\bibinfo {title} {Generalized gradient approximation made simple},\
  }\href {https://doi.org/10.1103/PhysRevLett.77.3865} {\bibfield  {journal}
  {\bibinfo  {journal} {Phys. Rev. Lett.}\ }\textbf {\bibinfo {volume} {77}},\
  \bibinfo {pages} {3865} (\bibinfo {year} {1996})}\BibitemShut {NoStop}%
\bibitem [{\citenamefont {Mostofi}\ \emph {et~al.}(2008)\citenamefont
  {Mostofi}, \citenamefont {Yates}, \citenamefont {Lee}, \citenamefont {Souza},
  \citenamefont {Vanderbilt},\ and\ \citenamefont {Marzari}}]{wannier}%
  \BibitemOpen
  \bibfield  {author} {\bibinfo {author} {\bibfnamefont {A.~A.}\ \bibnamefont
  {Mostofi}}, \bibinfo {author} {\bibfnamefont {J.~R.}\ \bibnamefont {Yates}},
  \bibinfo {author} {\bibfnamefont {Y.-S.}\ \bibnamefont {Lee}}, \bibinfo
  {author} {\bibfnamefont {I.}~\bibnamefont {Souza}}, \bibinfo {author}
  {\bibfnamefont {D.}~\bibnamefont {Vanderbilt}},\ and\ \bibinfo {author}
  {\bibfnamefont {N.}~\bibnamefont {Marzari}},\ }\bibfield  {title} {\bibinfo
  {title} {wannier90: A tool for obtaining maximally-localised wannier
  functions},\ }\href
  {https://doi.org/https://doi.org/10.1016/j.cpc.2007.11.016} {\bibfield
  {journal} {\bibinfo  {journal} {Computer Physics Communications}\ }\textbf
  {\bibinfo {volume} {178}},\ \bibinfo {pages} {685} (\bibinfo {year}
  {2008})}\BibitemShut {NoStop}%
\bibitem [{nom()}]{nomad}%
  \BibitemOpen
  \href@noop {} {\bibinfo {title} {The calculation data can be accessed via
  nomad under
  \url{https://dx.doi.org/10.17172/NOMAD/2024.12.19-2}.}}\BibitemShut {Stop}%
\end{thebibliography}%
\let\addcontentsline\oldaddcontentsline

\supplement{Supplemental Material for Van-Hove tuning of Fermi surface instabilities through compensated metallicity}

\subsection{Functional renormalization group results}

\begin{figure}[t]
  \centering
  \includegraphics[width = 0.4\textwidth]{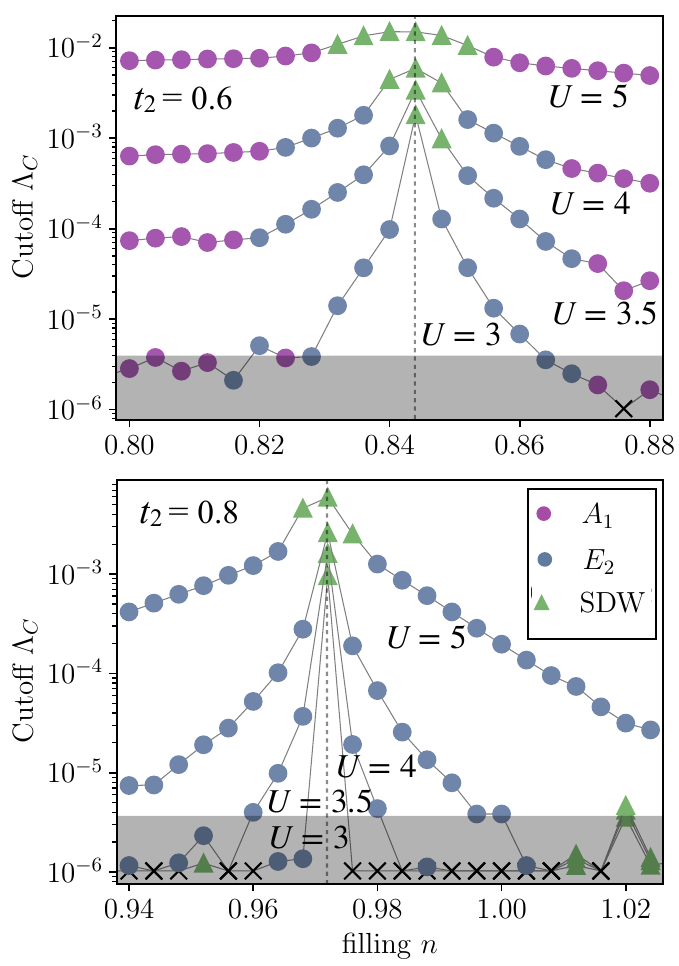}
  \caption{Critical cutoffs $\Lambda_C$ for the leading instabilities in the
           FRG as a function of onsite repulsion and doping around the lower
           vHs (dashed line). All values are given in units of $t_1$. The gray region indicates where the data becomes unreliable due to finite momentum resolution.
           The obtained results are consistent with the RPA results
           presented in Figure~2(b) of the main paper.
           }
  \label{fig:FRG}
\end{figure}

To substantiate our results obtained by large parameter space RPA calculations we
employ the functional renormalization group (FRG)
for two cross sections of our parameter space, Fig.~\ref{fig:FRG}.

The cross sections are taken for two fixed values of $t_2/t_1 = 0.6, 0.8$, varying the fillings
around the lower van Hove fillings of $n_{\text{lvH}} = 0.845, 0.972$ respectively.
The numerics were performed using an established FRG code \cite{Beyer2022}, which was benchmarked
against other FRG codes as \cite{Pena2017}. This code
utilizes the truncated unity approximation of FRG (TUFRG), which assumes that the
full two particle interaction vertex is well-approximated by interactions up to a
given cutoff range. In this case, the simulations included up fifth-nearest neighbors.
The interaction transfer momentum was discretized on a $36^2$ grid, with a further $51^2$
refinement used for the calculation of loops and loop derivatives.
In order to establish the stability of the phases, calculations
were run for several bare values of the on-site Hubbard repulsion $U$, from
approximately $1/3 < U/W < 1/2$, with bandwidth $W = 8.8 \, t_1$ for $t_2 = 0.6 \, t_1$
and $W = 10.5 \, t_1$ for $t_2 = 0.8 \, t_1$.

The results are in qualitative agreement with the RPA data presented in manuscript.
In accordance with Figure~2(a), the overall instability scale,
measured in the critical cutoff scale $\Lambda_C$, decreases as $t_2/t_1$ increases,
which we attribute to a reduced density of states around the Fermi level.

The vHs driven $E_2$ SC order is more persistent against a
detuning from perfect vH filling and the $A_1$ order is suppressed at
intermediate couplings.
This is a known drawback of the RPA procedure in the presence of small
pockets~\cite{Profe2024}:

  Due to the lack of cross-channel feedback, the RPA overestimates the
  contributions of ph fluctuations directly at the Fermi level, meaning features
  of RPA susceptibilities are heavily driven by the fermiology.
  In the manuscript, RPA sees a superconducting state in the $A_1$ irrep where
  the central pocket is small, circluar and has a large inverse Fermi velocity.
  In this case, the $A_1$ state maximally gaps out this density of states near
  the $\Gamma$ point.
  What FRG shows is that the van Hove physics can still dominate even if the
  van Hove singularity is near, but not directly on the Fermi surface, extending
  the range in filling of the $E_2$ SC state.

At higher interaction values, the vHs mediated ph fluctuations are directly
driving the system into a magnetic Stoner-like instability that is competing
with the $E_2$ SC order. Hence, at large $U$ and close to the pocket opening
($t_2/t_1 \sim 0.55$),
the $A_1$ SC phase eventually exceeds the $E_2$ state and occupies a large
part of the phase space away from vH filling.

\subsection{Strain induced charge compensation in group-IV 2D-"Xene" materials (X=Si, Ge, Sn)}


\begin{figure*}
    \centering
    \includegraphics[width = 0.97\textwidth]{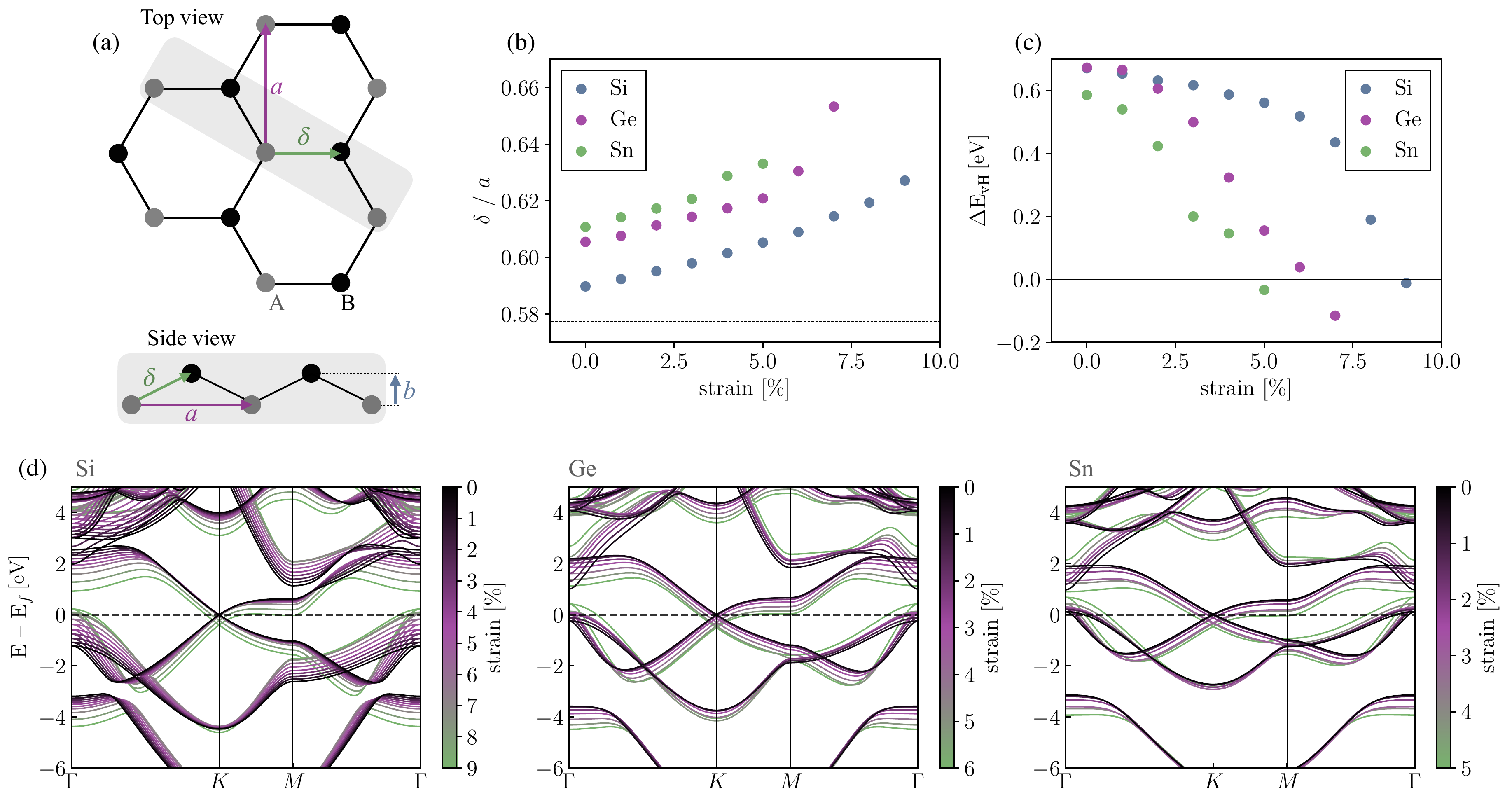}
    \caption{Xene lattice parameter with applied biaxial strain. (a) Real space
    position of the honeycomb sublattice sites in a buckled structure with
    (next) nearest neighbour distance $\delta$ ($a$). (b) Dependence of
    relative bond lengths for the Xene structures silicene, germanene and  stanene
    on strain. The strain is quantified as the enforced relative compression of
    the lattice constant of the fully relaxed structure. The black line
    indicates the $\delta / a$ ratio for the ideal honeycomb structure
    considered in the main text. (c) Energy offset of the upper van-Hove point
    from the Fermi level as a function of compressive strain for the different
    Xene compounds. (d)~Density functional theory band structures of silicene (Si), germanene (Ge), and stanene (St) for different values of strain. The dashed line indicates pristine half filling for each strain value individually.}
    \label{fig:1}
\end{figure*}

One promising route to achieve charge compensation in systems with honeycomb
structure is buckling. The bipartite unit cell allows for an out-of plane
displacement between sublattices without breaking the point group symmetry of
the honeycomb layer, as visualized in Fig.~\ref{fig:1}(a). By shrinking the
second nearest neighbour bond length \textit{w.r.t} the nearest neighbour distance,
this naturally increases the intra sublattice hybridization and facilitates the
charge compensation paradigm due to chiral symmetry breaking.

While the C atoms of graphene stabilize in an almost perfectly flat honeycomb
structure (except for potential rippling~\cite{Meyer2007}), its heavier relatives of the
main chemical group IV inherently acquire a buckled sublattice structure
facilitated by a $sp_3$-like hybridization (cf. Fig.~\ref{fig:1}(b)). This property,
however, impedes the synthesis of these materials in a free standing
configuration and necessitates a stabilizing substrate to sustain the
structure~\cite{Molle2017}.
Experimental and theoretical efforts to realize 2D layers of Silicon, Germanium, and Tin denoted as Xenes
(X=Si, Ge, Sn; \textit{i.e.} silicene, germanene, stanene) lead to a variety of
possible substrate hosts \cite{Zhu2015, Zhang2016, Li2020, DiSante2019, DiSante2019_2}.
As most of the works focus on maintaining and utilizing the topological
features concentrated around 2D Dirac cones,
our charge compensation paradigm reveals a novel avenue for experimental
investigations of Xene compounds: As demonstrated in many systems,
the lattice mismatch can be utilized to induce compressive (or tensile) strain
within the Xene structure, eliciting an increased (decreased) buckling~\cite{Molle2017, Wang2015, Gou2017,DiSante2019}.
Consequently, the intra-sublattice hybridization increases and facilitates
effective charge compensation under compressive strain. Previously, the resulting displacement of the Fermi level was considered as a disadvantage
as it inhibits the QSH insulator phase in favor of a metallic state.
However, the metallic phase enables
a topological SC transition, as proposed in this manuscript.

We conducted comprehensive DFT studies of silicene, germanene, and stanene
under compressive strain.
As displayed in Fig.~\ref{fig:1}(c), van-Hove filling at charge neutrality is
obtained for $5\, \%$ to $9 \, \%$ compression of the free standing lattice constant, which is in a order of magnitude proven to be experimentally accessible via epitaxial growth on \textit{e.g.} a
MoS$_2$ or Bi$_2$Te$_3$ substrate~\cite{Zhu2015, Zhang2016, Molle2017, DiSante2019}.
Deviating from the Honeycomb model employed in the main text, the Xene eigenspectra
in Fig.~\ref{fig:1}(d) feature an additional band stemming from
$sp_2$-derived hybrid orbitals, that eventually crosses the Fermi level close to $\Gamma$.
This leaves the vH singularity at pristine filling unaltered while providing
an additional hole pocket contributing in the charge compensation.
Thereby, the required long range hybridisation and hence strain for perfect
charge compensation at van-Hove is considerably reduced compared to the honeycomb
Hubbard model discussed in the main text.



\begin{figure}[!t]
    \centering
    \includegraphics[width = 0.4\textwidth]{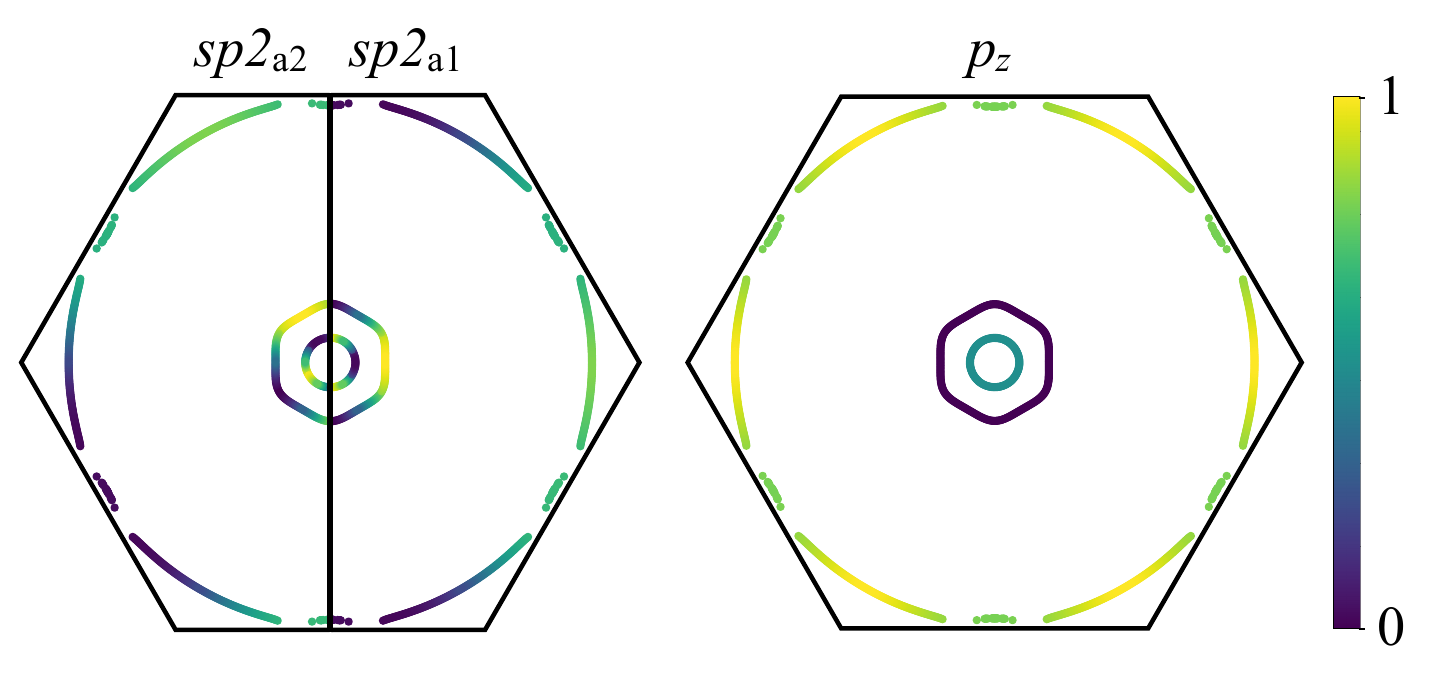}
    \caption{Orbital character on the FS for germanene strained by $6\%$. Two exemplary $sp_2$ orbitals are shown on the left. The $p_z$ orbital is mainly contributing to the outer Fermi pocket that is facilitating the $E_2$ SC phase. These properties are recovered qualitatively by silicene and stanene.}
    \label{fig:orbs}
\end{figure}

\subsection{Orbital selectivity and robust E$_2$ superconductivity in Xene compounds}
Analogous to the fermiology of the honeycomb Hubbard model at $t_2/t_1 \sim
0.85$, the strained Xene structures preserve the dominant FS nesting features of
the outer hexagonal Fermi pocket accountable for the vHs.
Despite the pronounced hexagonal shape of the additional hole-pocket, its
integrated density of states is small and we expect the vH pocket to be the main driver
for symmetry breaking transitions in the Xene compounds.
This suspicion is enhanced by an inspection of the orbital character of the FS
states in germanene: As evident from Fig.~\ref{fig:orbs}, the $p_z$ orbital is predominantly
contributing to the outer Fermi pocket.
Hence, the imbalance between intra and inter-pocket scattering from poor FS
nesting is further enhanced by not matching orbital makeup of the low lying
electronic states. This property is recovered qualitatively for the three Xene materials under investigation.

We take into account the multi-orbital nature of the Xene valence bands by
defining orbital dependent intra-orbital interactions.
Since the $sp_2$ orbitals are binding orbitals, that constitute the chemical bonds
of the crystal lattice, we expect larger screening of the Coulomb repulsion and
less impact on symmetry breaking transitions for these orbitals.
Hence, we set $U_{p_z} = 1.5\,$eV for the $p_z$ orbital and neglect the interaction within the $sp_2$ manifold.

Determining the superconducting phase for strained Xene structures at van-Hove
filling within the random phase approximation reveals robust $d$-wave gaps
transforming within the $E_2$ irrep. The order parameter in
Fig.~2 of the main paper resembles the main features of the charge compensated
honeycomb model: The gap function allows for a topological $d + id$ SC state,
that gaps out the outer Fermi pocket completely, while the two hole-pockets are
barely participating in the pairing process.

\subsection{Effect of spin orbit coupling}

As evident from our ab-initio analysis (cf. Fig.~\ref{fig:1}(d)), Xene structures consisting of heavier atoms require less strain to reach the charge compensated regime. This comes at the expense of having sizable spin-orbit coupling (SOC), therefore deviating from the paradigmatic example of Xenes, i.e. graphene.
Our relativistic density functional theory calculations indicate a gap opening at the Dirac points with increasing gap size for increasing SOC.
Our DFT calculations for unstrained silicene, germanene, and stanene reveal a gap at the Dirac points of $\approx 0.002\,$eV, $\approx 0.024\,$eV, and $\approx 0.077\,$eV, respectively.  Conversely, the vHs at the $M$-points remain spin degenerate.
For strained Xenes, however, Rashba-type SOC emerges due to the buckled structure \cite{gmitra2009band} which breaks inversion symmetry and would, in principle, allow for mixing of singlet and triplet superconducting states. It was recently shown that on hexagonal lattices such singlet-triplet mixing indeed occurs in the presence of Rashba SOC, but the $E_2$ irrep and chiral superconductivity prevail\,\cite{bunney2024chern}. Moreover, the emergent Rashba SOC induces a spin splitting at the vHs \cite{gmitra2009band} which, according to out DFT calculations, is negligible in the present case. As long as this gap is small, it was shown that the superconducting states nearby are not affected\,\cite{wolf2022topological}.



\subsection{Details of density functional theory calculations}
The DFT calculations were performed with the Vienna ab initio simulation
package (VASP) \cite{VASP} within the projector augmented-planewave (PAW)
method \cite{PAW1,PAW2}. For the exchange-correlation potential the PBE-GGA
functional \cite{PBE} was used, by expanding the Kohn-Sham wave functions into
plane waves up to an energy cutoff of 400 eV. We sample the Brillouin zone on a
12 × 12 × 1 Monkhorst-Pack mesh. We considered the free-standing buckled
monolayers silicene, germanene and stanene. To avoid unwanted interactions
between adjacent sheets a vacuum distance of 15A was employed. The structures
are relaxed in the out-of-plain direction until forces converged below 0.01
eV/Å. With Wannier90 8-orbital tight binding Hamiltonians were derived for
the $p_z$ and the $sp_2$ hybrid orbitals \cite{wannier}.
The input files of the density functional theory calculations are
publicly available~\cite{nomad}.

\end{document}